\documentclass[aps,reprint,pre,nofootinbib]{revtex4-2}
\usepackage{graphicx}
\usepackage{subfigure}
\usepackage[table,dvipsnames]{xcolor}
\usepackage{longtable}
\usepackage{multirow}
\usepackage{epstopdf, epsfig}
\usepackage{amsmath,esint}
\usepackage{amsfonts,amssymb}
\usepackage{dsfont}
\usepackage{dcolumn}
\usepackage{bm}
\usepackage{wasysym}
\usepackage[colorlinks=true, linkcolor=blue,citecolor=blue]{hyperref}

\setcitestyle{square}

\begin{document}

\title{Image rotation in plasmas}
\author{Renaud Gueroult}
\affiliation{LAPLACE, Universit{\'e} de Toulouse, CNRS, INPT, UPS, 31062 Toulouse, France}
\author{Shreekrishna K. Tripathi, Jia Han, Patrick Pribyl}
\affiliation{Department of Physics and Astronomy, University of California, Los Angeles, Los Angeles, California 90095, USA}
\author{Jean-Marcel Rax}
\affiliation{IJCLab, Universit{\'e} de Paris-Saclay, 91405 Orsay, France}
\author{Nathaniel J. Fisch}
\affiliation{Department of Astrophysical Sciences, Princeton University, Princeton, New Jersey 08540, USA}
\begin{abstract}
Because of the speed of light compared to material motion, the dragging of light is difficult to observe under laboratory conditions. Here we report on the first observation of image rotation, i.~e. a dragging by the medium of the wave's transverse structure, of Alfv{\'e}n waves in plasmas. Exploiting the naturally slow group velocity of these waves, significant wave rotation is achieved for modest angular frequency. Control over the rotation of the wave's structure is demonstrated through the plasma rotation imposed by biased electrodes. Remarkably, experimental results are well reproduced by light dragging theory derived for isotropic media, even if magnetized plasmas are anisotropic. In addition to offering new insights into the fundamental issue of angular momentum coupling between waves and media, these findings also open possibilities for new remote rotation sensing tools. 
\end{abstract}

\date{\today}

\maketitle

\emph{Introduction.--} Light dragging, as first recognized by Fresnel~\cite{Fresnel1818} and observed by Fizeau~\cite{Fizeau1851}, is one of the most influential concepts of physics in the nineteenth
century.  It is, together with a number of other fascinating manifestations including the Doppler effect, negative refraction~\cite{Grzegorczyk2006} or optical analogs to black holes' event horizon~\cite{Philbin2008}, evidence that wave propagation is modified by the medium’s motion. 

When the wavevector has a component perpendicular to the motion, the motion is the source of a deviation of the beam~\cite{Jones1972,Jones1975}. By analogy with Fresnel's original contribution on the effect of a longitudinal motion~\cite{Fresnel1818}, this effect is referred to as transverse Fresnel drag~\cite{Player1975,Carusotto2003}. For a rotation, transverse drag manifests as a rotation of the transverse structure of a wave - a phenomenon known as image rotation~\cite{Padgett2006}. Fundamentally, this arises from a coupling between the wave's orbital angular momentum (OAM) and the medium's angular momentum~\cite{Goette2007,Wisniewski-Barker2014}. This effect is analogous to the coupling with the wave's spin angular momentum (SAM) that manifests in the form of the polarization drag~\cite{Jones1976,Player1976}, as first conjectured by Thomson~\cite{Thomson1885} and Fermi~\cite{Fermi1923}. Angular momentum exchange between these channels can go either way, i. e. spin the medium or spin the wave. In that regard it bears similarities with the wave amplification in a rotating medium predicted by Zel’dovich~\cite{ZelDovich1971,ZelDovich1972}, and recently observed for OAM carrying sound waves~\cite{Cromb2020,Gooding2021}.

Observing light dragging effects experimentally is, however, difficult since they are typically very small in ordinary dielectrics~\cite{Jones1972,Jones1975,Jones1976}. One avenue to circumvent this issue that has been largely explored in the last decade is to use the artificially large group index~\cite{Franke-Arnold2011,Artoni2001,Safari2016,Kuan2016,Qin2020,Solomons2020} achieved in stimulated media (e.~g. through coherent population oscillation~\cite{Bigelow2003} or electromagnetically induced transparency~\cite{Hau1999,Fleischhauer2005}). Another route is to use the extremely high rotation that can be achieved in dilute media~\cite{Steinitz2020,Milner2021}. Here we show for the first time that a sizeable image rotation can in fact be observed for moderate rotation in natural media, namely plasmas, thanks to the inherently large group index of Alfv{\'e}n waves. Because a magnetized plasma is anisotropic, these results are also the first observation of image rotation in anisotropic media. Apart from the intrinsic academic interest in observing these fundamental phenomena, the ability to relate plasma rotation to image rotation is a useful two-way street: having demonstrated image rotation in rotating plasmas, one can now also deduce from image rotation plasma parameters including plasma rotation. Determining image rotation in plasmas is also important to quantify its effects on novel OAM-sensing tools designed to probe distant environments, such as the recent inference of the spin of the M87 black hole~\cite{Tamburini2011,Tamburini2022}. Observing image rotation finally informs on the opposite mechanism, in which waves can drive plasma rotation through momentum input~\cite{Ochs2021WaveDriven,Ochs2022MomentumConservation,Ochs2024Invited}.

\emph{Experimental setup.--} Experiments were conducted in the Large Plasma Device (LAPD) at the University of California, Los Angeles~\cite{Gekelman2016}. LAPD uses a large 38 cm diameter hot LaB6 cathode~\cite{Qian2023} to produce a highly reproducible 20 m long, 75 cm diameter magnetized plasma column with a repetition rate of $0.1-1$~Hz. Here we consider a dataset obtained in Helium with a background axial magnetic field $B_{0}=1$~kG, and a discharge time (calculated from plasma breakdown) of $15$~ms.  The average plasma density during the discharge is $n_{e}\sim5~10^{18}$~m$^{-3}$, with typically $20\%$ variation along the column. The corresponding reduced parameters are summarized in Tab.~\ref{Tab:conditions}.

\begin{table}
\begin{center}
\caption{\label{Tab:conditions}Main dimensional and reduced parameters for the experiment. Uncertainties come from axial density gradients.}
\begin{tabular}{c c c}
\hline
\hline
Magnetic field [kG] & $B_{0}$ & $1$\\ 
Plasma density [m$^{-3}$] & $n_{e}$ & $(5\pm1)~10^{18}$\\ 
Ion cyclotron freq. [kHz] & $\Omega_{ci}$  & $381$ \\
Alfv{\'en} velocity [m.s$^{-1}$]& $v_{A}$&  $(4.8\pm0.5)~10^{5}$\\
Wave to ion cycl. freq. ratio & $\omega_{0}/\Omega_{ci}$ & $0.4$\\
\hline
\hline\vspace{-1.1cm}
\end{tabular}
\end{center}
\end{table} 

\begin{figure*}
\begin{center}
\includegraphics[]{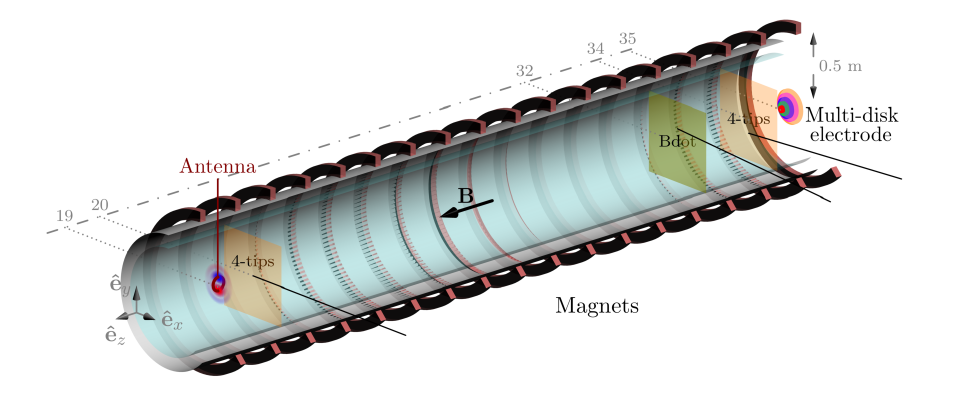}
\caption{Experimental setup in LAPD. Waves are generated by a single loop antenna on port $\#19$, $z=-6.3$~m. A multi-disk electrode (MDE) installed on port $\#35$ ($z=-11.66$~m) is used to control the plasma rotation. The wave and plasma parameters are diagnosed at three axial locations ($\#20$, $\#32$ and $\#34$) along the plasma column section. \vspace{-0.8cm}}
\label{Fig:Setup}
\end{center}
\end{figure*}

The particular setup used in this campaign is shown in Fig.~\ref{Fig:Setup}. It uses a $5$~m long section of the machine. 
On the left end of this section, which is the end closest to the plasma source, is positioned at $z_{a}=-6.3~$m a single loop antenna $10$~cm in diameter, with its normal along $\mathbf{\hat{e}}_{x}$, that is perpendicular to the background magnetic field $B_0\mathbf{\hat{e}}_{z}$. The antenna is slightly off axis, with $(x_{a},y_{a}) = (0.7,2.5)$~cm. It is powered via a variable frequency RF generator set at $f_{0}=154$~kHz, which corresponds to $\omega_{0}/\Omega_{ci}\sim0.4$ with $\Omega_{ci}=eB_{0}/m_{i}$ the ion cyclotron frequency.  The antenna creates a perpendicular magnetic perturbation $B_{x}\mathbf{\hat{e}}_{x}$, which in these conditions has been shown to couple to the plasma as a kinetic shear Alfv{\'e}n wave~\cite{Morales1997,Leneman1999,Gekelman2011} that propagates mostly parallel to the background magnetic field $B_{0}\mathbf{\hat{e}}_{z}$ (see End Matter). On the right end of this section is positioned a set of $5$ electrically insulated disk electrodes with radii $r_{i}=2.5i$ for $i\in[1,5]$ that are stacked concentrically. This multi-disk electrode (MDE) is centered and aligned on the machine axis. It has been demonstrated that by independently varying the electric bias on these electrodes, one can affect the plasma potential radial profile in LAPD, and from there cross-field plasma rotation in the plane normal to $\mathbf{B}_0$~\cite{Gueroult2024}. Plasma and wave parameters are measured as functions of time at different axial locations along the plasma column section. Specifically, probes are 
mounted on computer driven probe drives to acquire 2d maps of the floating potential on ports $\#20$ and $\#34$, that is respectively $33$~cm and $5$~m from the antenna ($5$~m and $33$~cm from the MDE). Concurrently, a three-axis differential magnetic probe (b-dot)~\cite{Everson2009} is mounted on another probe drive to acquire 2d maps of $d\mathbf{B}/dt$ on port $\#32$, that is $4.3$~m from the antenna ($1$~m from the MDE).

As summarized in Table~\ref{Tab:Biases}, we focus here on three specific biasing scenarios, which correspond to three different potential radial gradients imposed on the disk electrodes. We will verify that these scenarios lead to opposite radial electric fields in the plasma, and thus to both clockwise and counterclockwise rotation of the plasma column. In all cases the active biasing phase starts at $t=12~$ms, that is $3~$ms before the end of the main discharge. We will examine more specifically data at $t_{1}=11.95$~ms and $t_{2}=12.25~$ms, that is instants prior to and during biasing, respectively.

\begin{table}
\begin{center}
\caption{\label{Tab:Biases}Bias $\phi_{i}$ imposed on each of the $5$ disk electrodes for the three biasing scenario used in this study.}
\begin{tabular}{c c c c c c}
\hline
\hline
\multirow{2}{*}{Biasing scenario} & \multicolumn{5}{c}{Electrode bias [V]}\\
 & $\phi_{1}$ & $\phi_{2}$ & $\phi_{3}$ & $\phi_{4}$ & $\phi_{5}$\\
\hline
$\mathcal{A}$ - negative gradient & $30$ & $20$ & $15$ & $10$ & $5$\\
$\mathcal{B}$ - no bias & $-$ & $-$ & $-$ & $-$ & $-$\\
$\mathcal{C}$ - positive gradient & $0$ & $3$ & $30$ & $30$ & $7$\\
\hline
\hline\vspace{-1.cm}
\end{tabular}
\end{center}
\end{table}

\begin{figure}
\begin{center}
\includegraphics[]{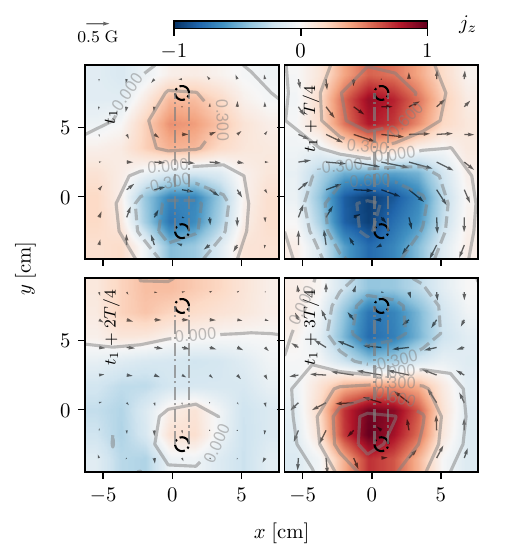}
\vspace{-0.4cm}\caption{Normalized axial current $j_{z}$ (colormap) and perpendicular wave magnetic field (arrows) in the $(xy)$ plane around the antenna center $(x_{a},y_{a}) = (0.7,2.5)$~cm on port $\#32$ at four successive instants (left to right, top to bottom) during one wave cycle $T=2\pi/\omega$ showing two antiparallel current filaments. Data is taken at at $t_{1}=11.95$~ms, that is just before bias is turned on. The dashed-dotted line indicates the antenna's position.\vspace{-0.8cm} }
\label{Fig:Wave_structure}
\end{center}
\end{figure}

\emph{Results.--} We first characterize the wave's transverse structure absent rotation. For this we examine the axial current $j_{z}$ associated with the wave, which is inferred by computing the axial component of $\bm{\nabla}\times\mathbf{B}$, and consider data before biasing has been turned on. Typical data obtained after numerically processing the data from the pickup probes on port $\#32$ with a bandpass filter $[0.5,1.5]f_{0}$ is plotted in Fig.~\ref{Fig:Wave_structure}, together with the perpendicular wave magnetic field whose amplitude is a fraction of a Gauss. The wave pattern features an azimuthal dependency $\exp{i\theta}$, with two antiparallel current filaments, which does not rotate with time~\cite{SupplementaryMaterial}. The orientation of the transverse structure depends very weakly on the biasing scenarios and frequency, with the symmetry axis mostly along $\mathbf{\hat{e}}_x$, consistent with an antenna along $\mathbf{\hat{e}}_y$. This structure is consistent with two waves with OAM $\pm\hbar$ propagating azimuthally in opposite directions, as documented by Rahbarnia\emph{~et al.}~\cite{Rahbarnia2010}

\begin{figure*}
\begin{center}
\includegraphics[]{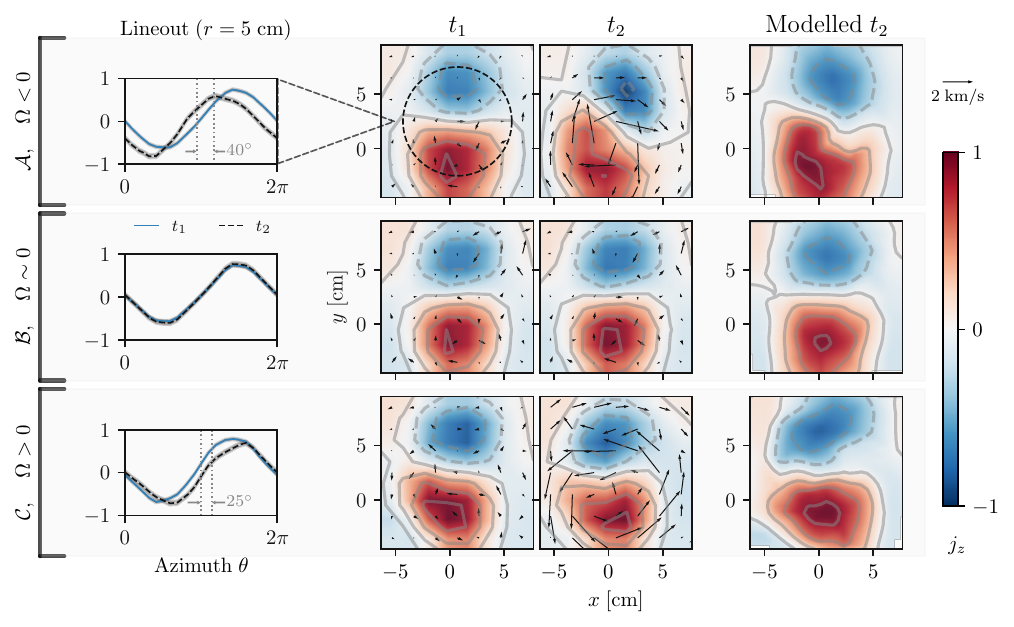}
\vspace{-0.4cm}\caption{Colormap of the normalized axial current $j_{z}$ in the $(xy)$ plane around the antenna center $(x_{a},y_{a}) = (0.7,2.5)$~cm on port $\#32$ at a given phase of a wave cycle before biasing ($t_{1}$, left column), during active biasing ($t_{2}$, middle column) and reconstructed from theory (right column). For the two leftmost columns the overlaid arrows show the velocity inferred from the floating potential at the same instants. The current density lineouts on the left, computed along the dotted circle shown on the top-left map, highlight the azimuthal shift of the current channels during biasing, with the gray shading indicating the less than $5\%$ variations measured over ten wave cycles. Each row corresponds to a different biasing scenario. \vspace{-0.8cm}}
\label{Fig:RotationDuringBiasing}
\end{center}
\end{figure*}

We can now examine how this transverse wave structure is affected by motion, and consider for this data at $t_{2}=12.25~$ms, i.~e. $250$~ms after biasing has been turned on. As seen going from first to second column in Fig.~\ref{Fig:RotationDuringBiasing}, we observe a rotation of the wave's transverse structure for the top and bottom rows, which corresponds to the active biasing scenarios $\mathcal{A}$ and $\mathcal{C}$. Conversely, the structure is seen to remain essentially the same for the no-bias scenario $\mathcal{B}$. This is confirmed when plotting an azimuthal lineout of the polar current density map computed with respect to the antenna's center, which highlights as shown on the left side of Fig.~\ref{Fig:RotationDuringBiasing} an azimuthal shift of the current channels during active biasing. As shown in Supplemental Material~\cite{SupplementaryMaterial}, this shift is verified to be stable over time, with azimuthal lineouts varying by less than $5\%$ over ten wave cycles. Trying to make sense of these results, we overlay with arrows on the two leftmost current maps in Fig.~\ref{Fig:RotationDuringBiasing} the velocity field inferred from the floating potential data taken at the same times on port $\#34$, considering only cross-field rotation. A first observation is that there is, other than for the no-bias scenario $\mathcal{B}$, a substantial change in flow between $t_{1}$ and $t_{2}$, consistent with the bias activation at $t=12~$ms. Quantitatively, velocities up to $3$ km.s$^{-1}$ are observed under biasing. Second, we verify that the negative gradient biasing scenario  $\mathcal{A}$, which corresponds to $\partial\phi/\partial r\leq0$, leads to a clockwise rotation $\Omega=(rB_{0})^{-1}\partial\phi/\partial r<0$, whereas the positive gradient biasing scenario  $\mathcal{C}$ yields $\Omega>0$.  Comparing finally the flow and the wave structure in the middle column, it appears that the wave structure is indeed dragged by the flow, with a rightward tilt on the top row, and a leftward tilt for the bottom row. Note that since the antenna is slightly off axis compared to the machine axis and the electrodes, one should not expect a pure rotation of the wave pattern, but rather a distortion.

\emph{Wave drag.--} To confirm this conjecture, we first evaluate the wave drag predicted by theory. Writing $\Sigma$ and $\Sigma'$ the laboratory frame and the plasma's rest-frame, and denoting with a prime quantities expressed in $\Sigma'$, we consider a wave propagating with group velocity $\mathbf{v_g}'=v_A \mathbf{\hat{e}}_z$ in the rest-frame, with $v_A$ the Alfv{\'e}n velocity. This is characteristic of an Alfv{\'e}n wave propagating parallel to $\mathbf{B}_{0}$, as it verifies the dispersion relation
\begin{equation}
\label{Eq:disp_Alf}
\omega' = k' v_A.
\end{equation}
For non-relativistic plasma velocities $\mathbf{v}$ and $v_A/c\ll 1$, a simple Galilean transformation then gives the lab-frame group velocity $\mathbf{v_g} = v_A \mathbf{\hat{e}}_z + \mathbf{v}$. For an observer at rest in $\Sigma$, plasma motion in the plane perpendicular to $\mathbf{B}_{0}$ thus introduces a perpendicular component to the wave's group velocity~\cite{Acheson1973}: the wave is being dragged by the medium. Specifically, the lab-frame group velocity is now tilted by an angle $v/v_g'=\beta n_g'$ with respect to the magnetic field direction, with $\beta=v/c$ the normalized medium's velocity and $n_{g}'$ the medium's rest-frame group index. After propagating a distance $l$, this tilt is the source of a perpendicular shift of the beam
\begin{equation}
\label{Eq:drag}
d\sim l v/v_{g}'\sim l v/v_{A}.
\end{equation}
This lateral shift $d$ is identical to the transverse drag known for isotropic dielectrics in the limit $n_g'\gg1$~\cite{Player1975,Jones1975}. It can be interpreted as the displacement of the medium during the time interval $l/v_{g}'$ over which light travels across it. Quantitatively, for the $30$~V voltage drop imposed on the electrodes across about $10~$cm, one gets a radial electric field of a few $100$~V.m$^{-1}$. It then gives a cross field velocity $v=E_r/B_0$ of a few km.s$^{-1}$, consistent with observations. For an average $v_{A}=480~$~km.s$^{-1}$ as given in Tab.~\ref{Tab:conditions} and a propagation length $l$ of a few meters, this gives $d\sim1-10$~cm, which appears consistent with the beam deviation seen in Fig.~\ref{Fig:RotationDuringBiasing}.

To substantiate further this analysis, we now compare the transverse wave structure measured during biasing with the prediction from theory when considering the effect of the non-uniform perpendicular velocity field created by biasing on the wave structure prior to biasing. Specifically, we compute the field 
\begin{equation}
\label{Eq:dragged_j_z}
j_{z}[\mathbf{x}+\bm{\delta}_{\mathbf{x}}(l)]= j_{z}(\mathbf{x},t_{1}),
\end{equation}
where we have introduced $\bm{\delta}_{\mathbf{x}}(z)$ the change in position in the perpendicular plane due to drag, for a ray starting at $\mathbf{x}=(x,y)$ in the plane of the antenna. This field is hence the image of the wave structure at the antenna but transported by the medium's motion, as it propagates along the column. From Eq.~\eqref{Eq:drag} the incremental drag writes
\begin{equation}
\frac{d\bm{\delta}_{\mathbf{x}}(z)}{dz} = \frac{\mathbf{v}[\mathbf{x}+\bm{\delta}_{\mathbf{x}}(z)]}{v_A}
\end{equation}
with $\mathbf{v}$ the perpendicular components of the cross-field velocity, which is inferred from the floating potential measurements on port $\#34$ at $t=t_{2}$. Results obtained for the parameters listed in Tab.~\ref{Tab:conditions} and $l=4.3$~m (distance between the antenna and the plane where the wave field is measured) for each of the three biasing scenarios are plotted in the right column of Fig.~\ref{Fig:RotationDuringBiasing}. We verify that the modelled dragged transverse structure indeed matches well the data measured at $t_{2}$ plotted in the middle column. It notably reproduces with good accuracy the amplitude of the clockwise and counter clockwise rotation of the current channels and of the zero current iso-contour, relative to the structure prior to biasing. This agreement is made more remarkable by the fact that the data for all three cases is obtained from the same simple formula Eq.~\eqref{Eq:dragged_j_z}, i. e. with no additional fitting parameter, and that axial density gradients are expected to lead to $10\%$ axial variations in $v_A$. Altogether this supports the analysis of the observed wave pattern shift as wave drag. 

Finally, to confirm the consistency of wave drag and image rotation, we consider these results in light of the recently proposed analytical model for image rotation for shear Alfv{\'en} wave~\cite{Rax2023b}. Although the angle of rotation of the transverse structure predicted in that case is more intricate since, because it considers OAM carrying wave, it does not rely on plane waves, one can show that the two results are consistent (see End Matter). Specifically, the rotation per unit length along $\mathbf{\hat{e}}_{z}$ is in this limit
\begin{equation}
\label{Eq:varphi}
\varphi = \frac{1}{2}\frac{\Omega}{\omega}k_{\parallel}\sim \frac{1}{2}\frac{\Omega}{v_A},
\end{equation}
which, given the trigonometric relation $r\varphi=d$ with $r=\sqrt{x^{2}+y^{2}}$ the radial coordinate and $v=r\Omega$, yields back other than for a factor one half the transverse drag scaling $d\sim lv/v_{A}$ identified in Eq.~\eqref{Eq:drag}.

\emph{Discussion.--} The results obtained above, showing remarkable agreement with observations, were obtained modeling the dispersion properties of the wave as a simple low frequency shear Alfv{\'e}n wave propagating exactly parallel to the confining magnetic field. Since $\omega_{0}/\Omega_{ci}=0.4$ in the experiment, and because a finite $k_{\perp}$ is expected for an antenna with finite transverse extent, these hypotheses are not strictly verified here. In these conditions the excited wave is in fact known to be a kinetic shear Alv{\'e}n wave (KSAW)~\cite{Leneman1999,Gekelman2011}, which is characterized by a more intricate dispersion relation than Eq.~\eqref{Eq:disp_Alf}. Yet, because the rest-frame group-velocity remains nearly parallel to the magnetic field (see End Matter), the wave drag predicted for KSAW is analogous to the one exposed for a simple Alfv{\'e}n wave. The main difference is a small enhancement of the wave drag angle due to a reduced parallel rest-frame group velocity. For the experimental parameters considered here, this drag increase is estimated to be about $15\%$. Implementing this correction addresses the slight underestimation of the pattern rotation in the rightmost column in Fig.~\ref{Fig:RotationDuringBiasing}, resulting in an improved agreement between theory and observations. 


The close agreement reported here between data and model is also of interest because image rotation scaling~\cite{Padgett2006,Goette2007,Franke-Arnold2011}, and the interpretation of polarization drag and image rotation as two manifestations of angular momentum coupling, were derived for isotropic media~\cite{Wisniewski-Barker2014}. In demonstrating now that these derivations hold also for the dispersive anisotropic media that are magnetized plasmas, we now broaden the demonstrated validity of these assertions, contributing importantly to the long-standing debate of how wave momentum manifests in media~~\cite{Pfeifer2007,Barnett2010,Bliokh2017}.

\emph{Opportunity for new rotation sensing schemes.--} A promising prospect for the image rotation identified here is the ability to remotely infer plasma rotation through a measure of the rotation angle $\varphi$ in Eq.~\eqref{Eq:varphi}. Specifically, while the potential of OAM-carrying beams to detect remotely rotation has already been demonstrated~\cite{Lavery2013}, it is typically done in reflection on a solid spinning target~\cite{Emile2020,Anderson2022,Ren2022}. As such this technique does not lend itself well to measure azimuthal or poloidal rotation profiles in rotating plasma columns or torii. In taking place in volume rather than at the interface, image rotation addresses this issue, while preserving the intrinsic advantage of OAM-based sensing, notably the possibility to use high azimuthal mode numbers to increase resolution. 

Such rotation sensing scheme may prove particularly useful in astrophysics. Indeed, it has been shown that OAM-carrying waves can be spontaneously generated in space through a number of processes, such as maser~\cite{Gray2014}, turbulence~\cite{Sanchez2011} or even free electrons in circular motion~\cite{Katoh2017}, offering opportunities to probe rotating environments using image rotation. More generally, OAM-sensing has been suggested for a number of astronomical and astrophysical applications~\cite{Harwit2003,Thide2007,Elias2008,Anzolin2008}. A notable example is the recent determination of the spin of the M87 black hole~\cite{Tamburini2022} using the dynamics of OAM beams around black holes~\cite{Tamburini2011}. Because, like polarization drag has been shown to affect polarimetry~\cite{Gueroult2019}, image rotation induced by a rotating plasma screen in front of a source will contribute to the overall OAM signature, determining image rotation is essential for data interpretation in these novel diagnostics. 

Apart from interest in sensing rotation in distant astrophysical phenomena, there could also be utility in sensing rotation and the associated plasma parameters in extremely hot laboratory plasma, such as fusion devices based on centrifugal confinement mechanisms~\cite{Lehnert1971}. Separate from image rotation, the rotation in these devices may be enabled by other waves, in fact even waves amplified by the free energy in the plasma produced by the nuclear fusion itself~\cite{Fisch1992,Fetterman2008,Fetterman2010,Kolmes2024VoltageDrop}.


\emph{Summary.--} Our work shows that image rotation, that is the rotation of the transverse structure of a wave due to the medium's rotation, whose observation has been so far limited to stimulated media exhibiting artificially large group index, can in fact be observed in plasmas in a variety of contexts. Using the unique capabilities of the Large Plasma Device at UCLA, we show that the transverse structure of an Alfv{\'e}n wave propagating along a magnetized plasma column can be rotated in either direction via the control on the radial electric field provided by biased end-electrodes positioned at one end of the column. The measured rotation is shown to match the deviation predicted from the transverse drag, i.e. the lateral shift, experienced by a wave propagating perpendicularly to the motion, computed for the wave's group velocity and the medium cross-field rotation. This result is a clear evidence of angular momentum coupling between a wave and a rotating medium. Apart from the academic interest in extending the theory of wave rotation to anisotropic media, the results here are important also for developing new means for inferring plasma rotation and related properties of plasma.  The diagnostic implications may be particularly valuable for diagnosing  very distant rotating plasma, such as of astrophysical origin, or very hot plasma, such as occurring in fusion concepts based on the vigorous rotation of plasma.

\section*{Acknowledgments} The authors would like to thank Thomas Look, Zoltan Lucky, Marvin Drandell, and Tai Ly for their technical support and assistance, and Julien Langlois and Aymeric Braud for constructive discussions. This work was supported by the French Agence Nationale de la Recherche (ANR), under grant ANR-21-CE30-0002 (project WaRP). This research was performed in the framework of the MagnetUS program at the Basic Plasma Science Facility at UCLA, which is funded by the United States Department of Energy and the National Science Foundation. NJF acknowledges support from DOE Grant No. DE-SC0016072.
RG gratefully acknowledges financial support for travel expenses from the International
Action Program from CNRS INSIS.

\section*{Data availability statement}
The data that support the findings of
this article are openly available~\cite{Zenodo2025}.

\section*{References}

\begin{thebibliography}{67}%
\makeatletter
\providecommand \@ifxundefined [1]{%
 \@ifx{#1\undefined}
}%
\providecommand \@ifnum [1]{%
 \ifnum #1\expandafter \@firstoftwo
 \else \expandafter \@secondoftwo
 \fi
}%
\providecommand \@ifx [1]{%
 \ifx #1\expandafter \@firstoftwo
 \else \expandafter \@secondoftwo
 \fi
}%
\providecommand \natexlab [1]{#1}%
\providecommand \enquote  [1]{``#1''}%
\providecommand \bibnamefont  [1]{#1}%
\providecommand \bibfnamefont [1]{#1}%
\providecommand \citenamefont [1]{#1}%
\providecommand \href@noop [0]{\@secondoftwo}%
\providecommand \href [0]{\begingroup \@sanitize@url \@href}%
\providecommand \@href[1]{\@@startlink{#1}\@@href}%
\providecommand \@@href[1]{\endgroup#1\@@endlink}%
\providecommand \@sanitize@url [0]{\catcode `\\12\catcode `\$12\catcode `\&12\catcode `\#12\catcode `\^12\catcode `\_12\catcode `\%12\relax}%
\providecommand \@@startlink[1]{}%
\providecommand \@@endlink[0]{}%
\providecommand \url  [0]{\begingroup\@sanitize@url \@url }%
\providecommand \@url [1]{\endgroup\@href {#1}{\urlprefix }}%
\providecommand \urlprefix  [0]{URL }%
\providecommand \Eprint [0]{\href }%
\providecommand \doibase [0]{https://doi.org/}%
\providecommand \selectlanguage [0]{\@gobble}%
\providecommand \bibinfo  [0]{\@secondoftwo}%
\providecommand \bibfield  [0]{\@secondoftwo}%
\providecommand \translation [1]{[#1]}%
\providecommand \BibitemOpen [0]{}%
\providecommand \bibitemStop [0]{}%
\providecommand \bibitemNoStop [0]{.\EOS\space}%
\providecommand \EOS [0]{\spacefactor3000\relax}%
\providecommand \BibitemShut  [1]{\csname bibitem#1\endcsname}%
\let\auto@bib@innerbib\@empty
\bibitem [{\citenamefont {Fresnel}(1818)}]{Fresnel1818}%
  \BibitemOpen
  \bibfield  {author} {\bibinfo {author} {\bibfnamefont {A.}~\bibnamefont {Fresnel}},\ }\bibfield  {title} {\bibinfo {title} {Lettre d'{A}ugustin {F}resnel {\`a} {F}rançois {A}rago sur l' influence du mouvement terrestre dans quelques ph{\'e}nom{\`e}nes d'optique},\ }\href {http://catalogue.bnf.fr/ark:/12148/cb343780820} {\bibfield  {journal} {\bibinfo  {journal} {Ann. Chim. Phys.}\ }\textbf {\bibinfo {volume} {9}},\ \bibinfo {pages} {57} (\bibinfo {year} {1818})}\BibitemShut {NoStop}%
\bibitem [{\citenamefont {Fizeau}(1851)}]{Fizeau1851}%
  \BibitemOpen
  \bibfield  {author} {\bibinfo {author} {\bibfnamefont {H.}~\bibnamefont {Fizeau}},\ }\bibfield  {title} {\bibinfo {title} {Sur les hypoth{\`e}ses relatives {\`a} l'{\'e}ther lumineux, et sur une exp{\'e}rience qui para{\^i}t d{\'e}montrer que le mouvement des corps change la vitesse avec laquelle la lumi{\`e}re se propage dans leur int{\'e}rieur},\ }\href {https://gallica.bnf.fr/ark:/12148/bpt6k29901/f351.image.r=Fizeau} {\bibfield  {journal} {\bibinfo  {journal} {C. R. Acad. Sci. Paris}\ }\textbf {\bibinfo {volume} {33}},\ \bibinfo {pages} {349} (\bibinfo {year} {1851})}\BibitemShut {NoStop}%
\bibitem [{\citenamefont {Grzegorczyk}\ and\ \citenamefont {Kong}(2006)}]{Grzegorczyk2006}%
  \BibitemOpen
  \bibfield  {author} {\bibinfo {author} {\bibfnamefont {T.~M.}\ \bibnamefont {Grzegorczyk}}\ and\ \bibinfo {author} {\bibfnamefont {J.~A.}\ \bibnamefont {Kong}},\ }\bibfield  {title} {\bibinfo {title} {Electrodynamics of moving media inducing positive and negative refraction},\ }\href {https://doi.org/10.1103/physrevb.74.033102} {\bibfield  {journal} {\bibinfo  {journal} {Phys. Rev. B}\ }\textbf {\bibinfo {volume} {74}},\ \bibinfo {pages} {033102} (\bibinfo {year} {2006})}\BibitemShut {NoStop}%
\bibitem [{\citenamefont {Philbin}\ \emph {et~al.}(2008)\citenamefont {Philbin}, \citenamefont {Kuklewicz}, \citenamefont {Robertson}, \citenamefont {Hill}, \citenamefont {K{\"o}nig},\ and\ \citenamefont {Leonhardt}}]{Philbin2008}%
  \BibitemOpen
  \bibfield  {author} {\bibinfo {author} {\bibfnamefont {T.~G.}\ \bibnamefont {Philbin}}, \bibinfo {author} {\bibfnamefont {C.}~\bibnamefont {Kuklewicz}}, \bibinfo {author} {\bibfnamefont {S.}~\bibnamefont {Robertson}}, \bibinfo {author} {\bibfnamefont {S.}~\bibnamefont {Hill}}, \bibinfo {author} {\bibfnamefont {F.}~\bibnamefont {K{\"o}nig}},\ and\ \bibinfo {author} {\bibfnamefont {U.}~\bibnamefont {Leonhardt}},\ }\bibfield  {title} {\bibinfo {title} {Fiber-optical analog of the event horizon},\ }\href {https://doi.org/10.1126/science.1153625} {\bibfield  {journal} {\bibinfo  {journal} {Science}\ }\textbf {\bibinfo {volume} {319}},\ \bibinfo {pages} {1367} (\bibinfo {year} {2008})}\BibitemShut {NoStop}%
\bibitem [{\citenamefont {Jones}(1972)}]{Jones1972}%
  \BibitemOpen
  \bibfield  {author} {\bibinfo {author} {\bibfnamefont {R.~V.}\ \bibnamefont {Jones}},\ }\bibfield  {title} {\bibinfo {title} {{\textquotesingle}{F}resnel aether drag{\textquotesingle} in a transversely moving medium},\ }\href {https://doi.org/10.1098/rspa.1972.0081} {\bibfield  {journal} {\bibinfo  {journal} {Proc. R. Soc. A}\ }\textbf {\bibinfo {volume} {328}},\ \bibinfo {pages} {337} (\bibinfo {year} {1972})}\BibitemShut {NoStop}%
\bibitem [{\citenamefont {Jones}(1975)}]{Jones1975}%
  \BibitemOpen
  \bibfield  {author} {\bibinfo {author} {\bibfnamefont {R.~V.}\ \bibnamefont {Jones}},\ }\bibfield  {title} {\bibinfo {title} {{\textquotesingle}{A}ether drag{\textquotesingle} in a transversely moving medium},\ }\href {https://doi.org/10.1098/rspa.1975.0141} {\bibfield  {journal} {\bibinfo  {journal} {Proc. R. Soc. A}\ }\textbf {\bibinfo {volume} {345}},\ \bibinfo {pages} {351} (\bibinfo {year} {1975})}\BibitemShut {NoStop}%
\bibitem [{\citenamefont {Player}(1975)}]{Player1975}%
  \BibitemOpen
  \bibfield  {author} {\bibinfo {author} {\bibfnamefont {M.~A.}\ \bibnamefont {Player}},\ }\bibfield  {title} {\bibinfo {title} {Dispersion and the transverse aether drag},\ }\href {https://doi.org/10.1098/rspa.1975.0139} {\bibfield  {journal} {\bibinfo  {journal} {Proc. R. Soc. A}\ }\textbf {\bibinfo {volume} {345}},\ \bibinfo {pages} {343} (\bibinfo {year} {1975})}\BibitemShut {NoStop}%
\bibitem [{\citenamefont {Carusotto}\ \emph {et~al.}(2003)\citenamefont {Carusotto}, \citenamefont {Artoni}, \citenamefont {Rocca},\ and\ \citenamefont {Bassani}}]{Carusotto2003}%
  \BibitemOpen
  \bibfield  {author} {\bibinfo {author} {\bibfnamefont {I.}~\bibnamefont {Carusotto}}, \bibinfo {author} {\bibfnamefont {M.}~\bibnamefont {Artoni}}, \bibinfo {author} {\bibfnamefont {G.~C.~L.}\ \bibnamefont {Rocca}},\ and\ \bibinfo {author} {\bibfnamefont {F.}~\bibnamefont {Bassani}},\ }\bibfield  {title} {\bibinfo {title} {Transverse {F}resnel-{F}izeau drag effects in strongly dispersive media},\ }\href {https://doi.org/10.1103/physreva.68.063819} {\bibfield  {journal} {\bibinfo  {journal} {Phys. Rev. A}\ }\textbf {\bibinfo {volume} {68}},\ \bibinfo {pages} {063819} (\bibinfo {year} {2003})}\BibitemShut {NoStop}%
\bibitem [{\citenamefont {Padgett}\ \emph {et~al.}(2006)\citenamefont {Padgett}, \citenamefont {Whyte}, \citenamefont {Girkin}, \citenamefont {Wright}, \citenamefont {Allen}, \citenamefont {\"{O}hberg},\ and\ \citenamefont {Barnett}}]{Padgett2006}%
  \BibitemOpen
  \bibfield  {author} {\bibinfo {author} {\bibfnamefont {M.}~\bibnamefont {Padgett}}, \bibinfo {author} {\bibfnamefont {G.}~\bibnamefont {Whyte}}, \bibinfo {author} {\bibfnamefont {J.}~\bibnamefont {Girkin}}, \bibinfo {author} {\bibfnamefont {A.}~\bibnamefont {Wright}}, \bibinfo {author} {\bibfnamefont {L.}~\bibnamefont {Allen}}, \bibinfo {author} {\bibfnamefont {P.}~\bibnamefont {\"{O}hberg}},\ and\ \bibinfo {author} {\bibfnamefont {S.~M.}\ \bibnamefont {Barnett}},\ }\bibfield  {title} {\bibinfo {title} {Polarization and image rotation induced by a rotating dielectric rod: an optical angular momentum interpretation},\ }\href {https://doi.org/10.1364/ol.31.002205} {\bibfield  {journal} {\bibinfo  {journal} {Opt. Lett.}\ }\textbf {\bibinfo {volume} {31}},\ \bibinfo {pages} {2205} (\bibinfo {year} {2006})}\BibitemShut {NoStop}%
\bibitem [{\citenamefont {G{\"o}tte}\ \emph {et~al.}(2007)\citenamefont {G{\"o}tte}, \citenamefont {Barnett},\ and\ \citenamefont {Padgett}}]{Goette2007}%
  \BibitemOpen
  \bibfield  {author} {\bibinfo {author} {\bibfnamefont {J.~B.}\ \bibnamefont {G{\"o}tte}}, \bibinfo {author} {\bibfnamefont {S.~M.}\ \bibnamefont {Barnett}},\ and\ \bibinfo {author} {\bibfnamefont {M.}~\bibnamefont {Padgett}},\ }\bibfield  {title} {\bibinfo {title} {On the dragging of light by a rotating medium},\ }\href {https://doi.org/10.1098/rspa.2007.1871} {\bibfield  {journal} {\bibinfo  {journal} {Proc. R. Soc. A}\ }\textbf {\bibinfo {volume} {463}},\ \bibinfo {pages} {2185} (\bibinfo {year} {2007})}\BibitemShut {NoStop}%
\bibitem [{\citenamefont {Wisniewski-Barker}\ \emph {et~al.}(2014)\citenamefont {Wisniewski-Barker}, \citenamefont {Gibson}, \citenamefont {Franke-Arnold}, \citenamefont {Boyd},\ and\ \citenamefont {Padgett}}]{Wisniewski-Barker2014}%
  \BibitemOpen
  \bibfield  {author} {\bibinfo {author} {\bibfnamefont {E.}~\bibnamefont {Wisniewski-Barker}}, \bibinfo {author} {\bibfnamefont {G.~M.}\ \bibnamefont {Gibson}}, \bibinfo {author} {\bibfnamefont {S.}~\bibnamefont {Franke-Arnold}}, \bibinfo {author} {\bibfnamefont {R.~W.}\ \bibnamefont {Boyd}},\ and\ \bibinfo {author} {\bibfnamefont {M.~J.}\ \bibnamefont {Padgett}},\ }\bibfield  {title} {\bibinfo {title} {Mechanical {F}araday effect for orbital angular momentum-carrying beams},\ }\href {https://doi.org/10.1364/OE.22.011690} {\bibfield  {journal} {\bibinfo  {journal} {Opt. Express}\ }\textbf {\bibinfo {volume} {22}},\ \bibinfo {pages} {11690} (\bibinfo {year} {2014})}\BibitemShut {NoStop}%
\bibitem [{\citenamefont {Jones}(1976)}]{Jones1976}%
  \BibitemOpen
  \bibfield  {author} {\bibinfo {author} {\bibfnamefont {R.~V.}\ \bibnamefont {Jones}},\ }\bibfield  {title} {\bibinfo {title} {Rotary aether drag},\ }\href {https://doi.org/10.1098/rspa.1976.0082} {\bibfield  {journal} {\bibinfo  {journal} {Proc. R. Soc. A}\ }\textbf {\bibinfo {volume} {349}},\ \bibinfo {pages} {423} (\bibinfo {year} {1976})}\BibitemShut {NoStop}%
\bibitem [{\citenamefont {Player}(1976)}]{Player1976}%
  \BibitemOpen
  \bibfield  {author} {\bibinfo {author} {\bibfnamefont {M.~A.}\ \bibnamefont {Player}},\ }\bibfield  {title} {\bibinfo {title} {On the dragging of the plane of polarization of light propagating in a rotating medium},\ }\href {https://doi.org/10.1098/rspa.1976.0083} {\bibfield  {journal} {\bibinfo  {journal} {Proc. R. Soc. A}\ }\textbf {\bibinfo {volume} {349}},\ \bibinfo {pages} {441} (\bibinfo {year} {1976})}\BibitemShut {NoStop}%
\bibitem [{\citenamefont {Thomson}(1885)}]{Thomson1885}%
  \BibitemOpen
  \bibfield  {author} {\bibinfo {author} {\bibfnamefont {J.~J.}\ \bibnamefont {Thomson}},\ }\bibfield  {title} {\bibinfo {title} {Note on the rotation of the plane of polarization of light by a moving medium},\ }\href {https://www.biodiversitylibrary.org/page/26955928} {\bibfield  {journal} {\bibinfo  {journal} {Proc. Camb. Phil. Soc.}\ }\textbf {\bibinfo {volume} {5}},\ \bibinfo {pages} {250} (\bibinfo {year} {1885})}\BibitemShut {NoStop}%
\bibitem [{\citenamefont {Fermi}(1923)}]{Fermi1923}%
  \BibitemOpen
  \bibfield  {author} {\bibinfo {author} {\bibfnamefont {E.}~\bibnamefont {Fermi}},\ }\bibfield  {title} {\bibinfo {title} {Sul trascinamento del piano di polarizzazione da parte di un messo rotante},\ }\href {http://operedigitali.lincei.it/rendicontiFMN/rol/pdf/S5V32T1A1923P115_118.pdf} {\bibfield  {journal} {\bibinfo  {journal} {Rend. Mat. Acc. Lincei}\ }\textbf {\bibinfo {volume} {32}},\ \bibinfo {pages} {115} (\bibinfo {year} {1923})},\ \bibinfo {note} {reprinted in Collected Papers, vol. 1 (University of Chicago Press, Chicago, 1962)}\BibitemShut {NoStop}%
\bibitem [{\citenamefont {Zel'Dovich}(1971)}]{ZelDovich1971}%
  \BibitemOpen
  \bibfield  {author} {\bibinfo {author} {\bibfnamefont {Y.~B.}\ \bibnamefont {Zel'Dovich}},\ }\bibfield  {title} {\bibinfo {title} {Generation of waves by a rotating body},\ }\href {https://ui.adsabs.harvard.edu/abs/1971JETPL..14..180Z} {\bibfield  {journal} {\bibinfo  {journal} {JETP}\ }\textbf {\bibinfo {volume} {14}},\ \bibinfo {pages} {180} (\bibinfo {year} {1971})}\BibitemShut {NoStop}%
\bibitem [{\citenamefont {Zel'Dovich}(1972)}]{ZelDovich1972}%
  \BibitemOpen
  \bibfield  {author} {\bibinfo {author} {\bibfnamefont {Y.~B.}\ \bibnamefont {Zel'Dovich}},\ }\bibfield  {title} {\bibinfo {title} {Amplification of cylindrical electromagnetic waves reflected from a rotating body},\ }\href {https://ui.adsabs.harvard.edu/abs/1972JETP...35.1085Z} {\bibfield  {journal} {\bibinfo  {journal} {JETP}\ }\textbf {\bibinfo {volume} {35}},\ \bibinfo {pages} {1085} (\bibinfo {year} {1972})}\BibitemShut {NoStop}%
\bibitem [{\citenamefont {Cromb}\ \emph {et~al.}(2020)\citenamefont {Cromb}, \citenamefont {Gibson}, \citenamefont {Toninelli}, \citenamefont {Padgett}, \citenamefont {Wright},\ and\ \citenamefont {Faccio}}]{Cromb2020}%
  \BibitemOpen
  \bibfield  {author} {\bibinfo {author} {\bibfnamefont {M.}~\bibnamefont {Cromb}}, \bibinfo {author} {\bibfnamefont {G.~M.}\ \bibnamefont {Gibson}}, \bibinfo {author} {\bibfnamefont {E.}~\bibnamefont {Toninelli}}, \bibinfo {author} {\bibfnamefont {M.~J.}\ \bibnamefont {Padgett}}, \bibinfo {author} {\bibfnamefont {E.~M.}\ \bibnamefont {Wright}},\ and\ \bibinfo {author} {\bibfnamefont {D.}~\bibnamefont {Faccio}},\ }\bibfield  {title} {\bibinfo {title} {Amplification of waves from a rotating body},\ }\href {https://doi.org/10.1038/s41567-020-0944-3} {\bibfield  {journal} {\bibinfo  {journal} {Nat. Phys.}\ }\textbf {\bibinfo {volume} {16}},\ \bibinfo {pages} {1069} (\bibinfo {year} {2020})}\BibitemShut {NoStop}%
\bibitem [{\citenamefont {Gooding}\ \emph {et~al.}(2021)\citenamefont {Gooding}, \citenamefont {Weinfurtner},\ and\ \citenamefont {Unruh}}]{Gooding2021}%
  \BibitemOpen
  \bibfield  {author} {\bibinfo {author} {\bibfnamefont {C.}~\bibnamefont {Gooding}}, \bibinfo {author} {\bibfnamefont {S.}~\bibnamefont {Weinfurtner}},\ and\ \bibinfo {author} {\bibfnamefont {W.~G.}\ \bibnamefont {Unruh}},\ }\bibfield  {title} {\bibinfo {title} {Superradiant scattering of orbital angular momentum beams},\ }\href {https://doi.org/10.1103/physrevresearch.3.023242} {\bibfield  {journal} {\bibinfo  {journal} {Phys. Rev. Research}\ }\textbf {\bibinfo {volume} {3}},\ \bibinfo {pages} {023242} (\bibinfo {year} {2021})}\BibitemShut {NoStop}%
\bibitem [{\citenamefont {Franke-Arnold}\ \emph {et~al.}(2011)\citenamefont {Franke-Arnold}, \citenamefont {Gibson}, \citenamefont {Boyd},\ and\ \citenamefont {Padgett}}]{Franke-Arnold2011}%
  \BibitemOpen
  \bibfield  {author} {\bibinfo {author} {\bibfnamefont {S.}~\bibnamefont {Franke-Arnold}}, \bibinfo {author} {\bibfnamefont {G.}~\bibnamefont {Gibson}}, \bibinfo {author} {\bibfnamefont {R.~W.}\ \bibnamefont {Boyd}},\ and\ \bibinfo {author} {\bibfnamefont {M.~J.}\ \bibnamefont {Padgett}},\ }\bibfield  {title} {\bibinfo {title} {Rotary photon drag enhanced by a slow-light medium},\ }\href {https://doi.org/10.1126/science.1203984} {\bibfield  {journal} {\bibinfo  {journal} {Science}\ }\textbf {\bibinfo {volume} {333}},\ \bibinfo {pages} {65} (\bibinfo {year} {2011})}\BibitemShut {NoStop}%
\bibitem [{\citenamefont {Artoni}\ \emph {et~al.}(2001)\citenamefont {Artoni}, \citenamefont {Carusotto}, \citenamefont {La~Rocca},\ and\ \citenamefont {Bassani}}]{Artoni2001}%
  \BibitemOpen
  \bibfield  {author} {\bibinfo {author} {\bibfnamefont {M.}~\bibnamefont {Artoni}}, \bibinfo {author} {\bibfnamefont {I.}~\bibnamefont {Carusotto}}, \bibinfo {author} {\bibfnamefont {G.~C.}\ \bibnamefont {La~Rocca}},\ and\ \bibinfo {author} {\bibfnamefont {F.}~\bibnamefont {Bassani}},\ }\bibfield  {title} {\bibinfo {title} {Fresnel light drag in a coherently driven moving medium},\ }\href {https://doi.org/10.1103/PhysRevLett.86.2549} {\bibfield  {journal} {\bibinfo  {journal} {Phys. Rev. Lett.}\ }\textbf {\bibinfo {volume} {86}},\ \bibinfo {pages} {2549} (\bibinfo {year} {2001})}\BibitemShut {NoStop}%
\bibitem [{\citenamefont {Safari}\ \emph {et~al.}(2016)\citenamefont {Safari}, \citenamefont {De~Leon}, \citenamefont {Mirhosseini}, \citenamefont {Maga{\~n}a-Loaiza},\ and\ \citenamefont {Boyd}}]{Safari2016}%
  \BibitemOpen
  \bibfield  {author} {\bibinfo {author} {\bibfnamefont {A.}~\bibnamefont {Safari}}, \bibinfo {author} {\bibfnamefont {I.}~\bibnamefont {De~Leon}}, \bibinfo {author} {\bibfnamefont {M.}~\bibnamefont {Mirhosseini}}, \bibinfo {author} {\bibfnamefont {O.~S.}\ \bibnamefont {Maga{\~n}a-Loaiza}},\ and\ \bibinfo {author} {\bibfnamefont {R.~W.}\ \bibnamefont {Boyd}},\ }\bibfield  {title} {\bibinfo {title} {Light-drag enhancement by a highly dispersive rubidium vapor},\ }\href {https://doi.org/10.1103/PhysRevLett.116.013601} {\bibfield  {journal} {\bibinfo  {journal} {Phys. Rev. Lett.}\ }\textbf {\bibinfo {volume} {116}},\ \bibinfo {eid} {013601} (\bibinfo {year} {2016})}\BibitemShut {NoStop}%
\bibitem [{\citenamefont {Kuan}\ \emph {et~al.}(2016)\citenamefont {Kuan}, \citenamefont {Huang}, \citenamefont {Chan}, \citenamefont {Kosen},\ and\ \citenamefont {Lan}}]{Kuan2016}%
  \BibitemOpen
  \bibfield  {author} {\bibinfo {author} {\bibfnamefont {P.-C.}\ \bibnamefont {Kuan}}, \bibinfo {author} {\bibfnamefont {C.}~\bibnamefont {Huang}}, \bibinfo {author} {\bibfnamefont {W.~S.}\ \bibnamefont {Chan}}, \bibinfo {author} {\bibfnamefont {S.}~\bibnamefont {Kosen}},\ and\ \bibinfo {author} {\bibfnamefont {S.-Y.}\ \bibnamefont {Lan}},\ }\bibfield  {title} {\bibinfo {title} {Large fizeau’s light-dragging effect in a moving electromagnetically induced transparent medium},\ }\href {https://doi.org/10.1038/ncomms13030} {\bibfield  {journal} {\bibinfo  {journal} {Nat. Commun.}\ }\textbf {\bibinfo {volume} {7}},\ \bibinfo {pages} {13030} (\bibinfo {year} {2016})}\BibitemShut {NoStop}%
\bibitem [{\citenamefont {Qin}\ \emph {et~al.}(2020)\citenamefont {Qin}, \citenamefont {Yang}, \citenamefont {Zhang}, \citenamefont {Chen}, \citenamefont {Shen}, \citenamefont {Liu}, \citenamefont {Chen}, \citenamefont {Jiang}, \citenamefont {Chen},\ and\ \citenamefont {Wan}}]{Qin2020}%
  \BibitemOpen
  \bibfield  {author} {\bibinfo {author} {\bibfnamefont {T.}~\bibnamefont {Qin}}, \bibinfo {author} {\bibfnamefont {J.}~\bibnamefont {Yang}}, \bibinfo {author} {\bibfnamefont {F.}~\bibnamefont {Zhang}}, \bibinfo {author} {\bibfnamefont {Y.}~\bibnamefont {Chen}}, \bibinfo {author} {\bibfnamefont {D.}~\bibnamefont {Shen}}, \bibinfo {author} {\bibfnamefont {W.}~\bibnamefont {Liu}}, \bibinfo {author} {\bibfnamefont {L.}~\bibnamefont {Chen}}, \bibinfo {author} {\bibfnamefont {X.}~\bibnamefont {Jiang}}, \bibinfo {author} {\bibfnamefont {X.}~\bibnamefont {Chen}},\ and\ \bibinfo {author} {\bibfnamefont {W.}~\bibnamefont {Wan}},\ }\bibfield  {title} {\bibinfo {title} {Fast- and slow-light-enhanced light drag in a moving microcavity},\ }\href {https://doi.org/10.1038/s42005-020-0386-3} {\bibfield  {journal} {\bibinfo  {journal} {Commun. Phys.}\ }\textbf {\bibinfo {volume} {3}},\ \bibinfo {pages} {118} (\bibinfo {year} {2020})}\BibitemShut {NoStop}%
\bibitem [{\citenamefont {Solomons}\ \emph {et~al.}(2020)\citenamefont {Solomons}, \citenamefont {Banerjee}, \citenamefont {Smartsev}, \citenamefont {Friedman}, \citenamefont {Eger}, \citenamefont {Firstenberg},\ and\ \citenamefont {Davidson}}]{Solomons2020}%
  \BibitemOpen
  \bibfield  {author} {\bibinfo {author} {\bibfnamefont {Y.}~\bibnamefont {Solomons}}, \bibinfo {author} {\bibfnamefont {C.}~\bibnamefont {Banerjee}}, \bibinfo {author} {\bibfnamefont {S.}~\bibnamefont {Smartsev}}, \bibinfo {author} {\bibfnamefont {J.}~\bibnamefont {Friedman}}, \bibinfo {author} {\bibfnamefont {D.}~\bibnamefont {Eger}}, \bibinfo {author} {\bibfnamefont {O.}~\bibnamefont {Firstenberg}},\ and\ \bibinfo {author} {\bibfnamefont {N.}~\bibnamefont {Davidson}},\ }\bibfield  {title} {\bibinfo {title} {Transverse drag of slow light in moving atomic vapor},\ }\href {https://doi.org/10.1364/OL.394389} {\bibfield  {journal} {\bibinfo  {journal} {Opt. Lett.}\ }\textbf {\bibinfo {volume} {45}},\ \bibinfo {pages} {3431} (\bibinfo {year} {2020})}\BibitemShut {NoStop}%
\bibitem [{\citenamefont {Bigelow}\ \emph {et~al.}(2003)\citenamefont {Bigelow}, \citenamefont {Lepeshkin},\ and\ \citenamefont {Boyd}}]{Bigelow2003}%
  \BibitemOpen
  \bibfield  {author} {\bibinfo {author} {\bibfnamefont {M.~S.}\ \bibnamefont {Bigelow}}, \bibinfo {author} {\bibfnamefont {N.~N.}\ \bibnamefont {Lepeshkin}},\ and\ \bibinfo {author} {\bibfnamefont {R.~W.}\ \bibnamefont {Boyd}},\ }\bibfield  {title} {\bibinfo {title} {Observation of ultraslow light propagation in a ruby crystal at room temperature},\ }\href {https://doi.org/10.1103/PhysRevLett.90.113903} {\bibfield  {journal} {\bibinfo  {journal} {Phys. Rev. Lett.}\ }\textbf {\bibinfo {volume} {90}},\ \bibinfo {pages} {113903} (\bibinfo {year} {2003})}\BibitemShut {NoStop}%
\bibitem [{\citenamefont {Hau}\ \emph {et~al.}(1999)\citenamefont {Hau}, \citenamefont {Harris}, \citenamefont {Dutton},\ and\ \citenamefont {Behroozi}}]{Hau1999}%
  \BibitemOpen
  \bibfield  {author} {\bibinfo {author} {\bibfnamefont {L.~V.}\ \bibnamefont {Hau}}, \bibinfo {author} {\bibfnamefont {S.~E.}\ \bibnamefont {Harris}}, \bibinfo {author} {\bibfnamefont {Z.}~\bibnamefont {Dutton}},\ and\ \bibinfo {author} {\bibfnamefont {C.~H.}\ \bibnamefont {Behroozi}},\ }\bibfield  {title} {\bibinfo {title} {Light speed reduction to 17 metres per second in an ultracold atomic},\ }\href {https://doi.org/10.1038/17561} {\bibfield  {journal} {\bibinfo  {journal} {Nature}\ }\textbf {\bibinfo {volume} {397}},\ \bibinfo {pages} {594} (\bibinfo {year} {1999})}\BibitemShut {NoStop}%
\bibitem [{\citenamefont {Fleischhauer}\ \emph {et~al.}(2005)\citenamefont {Fleischhauer}, \citenamefont {Imamoglu},\ and\ \citenamefont {Marangos}}]{Fleischhauer2005}%
  \BibitemOpen
  \bibfield  {author} {\bibinfo {author} {\bibfnamefont {M.}~\bibnamefont {Fleischhauer}}, \bibinfo {author} {\bibfnamefont {A.}~\bibnamefont {Imamoglu}},\ and\ \bibinfo {author} {\bibfnamefont {J.~P.}\ \bibnamefont {Marangos}},\ }\bibfield  {title} {\bibinfo {title} {Electromagnetically induced transparency: Optics in coherent media},\ }\href {https://doi.org/10.1103/revmodphys.77.633} {\bibfield  {journal} {\bibinfo  {journal} {Rev. Modern Phys.}\ }\textbf {\bibinfo {volume} {77}},\ \bibinfo {pages} {633} (\bibinfo {year} {2005})}\BibitemShut {NoStop}%
\bibitem [{\citenamefont {Steinitz}\ and\ \citenamefont {Averbukh}(2020)}]{Steinitz2020}%
  \BibitemOpen
  \bibfield  {author} {\bibinfo {author} {\bibfnamefont {U.}~\bibnamefont {Steinitz}}\ and\ \bibinfo {author} {\bibfnamefont {I.~S.}\ \bibnamefont {Averbukh}},\ }\bibfield  {title} {\bibinfo {title} {Giant polarization drag in a gas of molecular super-rotors},\ }\href {https://doi.org/10.1103/physreva.101.021404} {\bibfield  {journal} {\bibinfo  {journal} {Phys. Rev. A}\ }\textbf {\bibinfo {volume} {101}},\ \bibinfo {pages} {021404} (\bibinfo {year} {2020})}\BibitemShut {NoStop}%
\bibitem [{\citenamefont {Milner}\ \emph {et~al.}(2021)\citenamefont {Milner}, \citenamefont {Steinitz}, \citenamefont {Averbukh},\ and\ \citenamefont {Milner}}]{Milner2021}%
  \BibitemOpen
  \bibfield  {author} {\bibinfo {author} {\bibfnamefont {A.~A.}\ \bibnamefont {Milner}}, \bibinfo {author} {\bibfnamefont {U.}~\bibnamefont {Steinitz}}, \bibinfo {author} {\bibfnamefont {I.~S.}\ \bibnamefont {Averbukh}},\ and\ \bibinfo {author} {\bibfnamefont {V.}~\bibnamefont {Milner}},\ }\bibfield  {title} {\bibinfo {title} {Observation of mechanical {F}araday effect in gaseous media},\ }\href {https://doi.org/10.1103/physrevlett.127.073901} {\bibfield  {journal} {\bibinfo  {journal} {Phys. Rev. Lett.}\ }\textbf {\bibinfo {volume} {127}},\ \bibinfo {pages} {073901} (\bibinfo {year} {2021})}\BibitemShut {NoStop}%
\bibitem [{\citenamefont {Tamburini}\ \emph {et~al.}(2011)\citenamefont {Tamburini}, \citenamefont {Thidé}, \citenamefont {Molina-Terriza},\ and\ \citenamefont {Anzolin}}]{Tamburini2011}%
  \BibitemOpen
  \bibfield  {author} {\bibinfo {author} {\bibfnamefont {F.}~\bibnamefont {Tamburini}}, \bibinfo {author} {\bibfnamefont {B.}~\bibnamefont {Thidé}}, \bibinfo {author} {\bibfnamefont {G.}~\bibnamefont {Molina-Terriza}},\ and\ \bibinfo {author} {\bibfnamefont {G.}~\bibnamefont {Anzolin}},\ }\bibfield  {title} {\bibinfo {title} {Twisting of light around rotating black holes},\ }\href {https://doi.org/10.1038/nphys1907} {\bibfield  {journal} {\bibinfo  {journal} {Nat. Phys.}\ }\textbf {\bibinfo {volume} {7}},\ \bibinfo {pages} {195} (\bibinfo {year} {2011})}\BibitemShut {NoStop}%
\bibitem [{\citenamefont {Tamburini}\ \emph {et~al.}(2022)\citenamefont {Tamburini}, \citenamefont {Feleppa},\ and\ \citenamefont {Thidé}}]{Tamburini2022}%
  \BibitemOpen
  \bibfield  {author} {\bibinfo {author} {\bibfnamefont {F.}~\bibnamefont {Tamburini}}, \bibinfo {author} {\bibfnamefont {F.}~\bibnamefont {Feleppa}},\ and\ \bibinfo {author} {\bibfnamefont {B.}~\bibnamefont {Thidé}},\ }\bibfield  {title} {\bibinfo {title} {Constraining the generalized uncertainty principle with the light twisted by rotating black holes and m87*},\ }\href {https://doi.org/10.1016/j.physletb.2022.136894} {\bibfield  {journal} {\bibinfo  {journal} {Phys. Lett. B}\ }\textbf {\bibinfo {volume} {826}},\ \bibinfo {pages} {136894} (\bibinfo {year} {2022})}\BibitemShut {NoStop}%
\bibitem [{\citenamefont {Ochs}\ and\ \citenamefont {Fisch}(2021)}]{Ochs2021WaveDriven}%
  \BibitemOpen
  \bibfield  {author} {\bibinfo {author} {\bibfnamefont {I.~E.}\ \bibnamefont {Ochs}}\ and\ \bibinfo {author} {\bibfnamefont {N.~J.}\ \bibnamefont {Fisch}},\ }\bibfield  {title} {\bibinfo {title} {Wave-driven torques to drive current and rotation},\ }\href {https://doi.org/10.1063/5.0062034} {\bibfield  {journal} {\bibinfo  {journal} {Phys. Plasmas}\ }\textbf {\bibinfo {volume} {28}},\ \bibinfo {pages} {102506} (\bibinfo {year} {2021})}\BibitemShut {NoStop}%
\bibitem [{\citenamefont {Ochs}\ and\ \citenamefont {Fisch}(2022)}]{Ochs2022MomentumConservation}%
  \BibitemOpen
  \bibfield  {author} {\bibinfo {author} {\bibfnamefont {I.~E.}\ \bibnamefont {Ochs}}\ and\ \bibinfo {author} {\bibfnamefont {N.~J.}\ \bibnamefont {Fisch}},\ }\bibfield  {title} {\bibinfo {title} {Momentum conservation in current drive and alpha-channeling-mediated rotation drive},\ }\href {https://doi.org/10.1063/5.0085821} {\bibfield  {journal} {\bibinfo  {journal} {Phys. Plasmas}\ }\textbf {\bibinfo {volume} {29}},\ \bibinfo {pages} {062106} (\bibinfo {year} {2022})}\BibitemShut {NoStop}%
\bibitem [{\citenamefont {Ochs}(2024)}]{Ochs2024Invited}%
  \BibitemOpen
  \bibfield  {author} {\bibinfo {author} {\bibfnamefont {I.~E.}\ \bibnamefont {Ochs}},\ }\bibfield  {title} {\bibinfo {title} {When do waves drive plasma flows?},\ }\href {https://doi.org/10.1063/5.0201780} {\bibfield  {journal} {\bibinfo  {journal} {Phys. Plasmas}\ }\textbf {\bibinfo {volume} {31}},\ \bibinfo {pages} {042116} (\bibinfo {year} {2024})}\BibitemShut {NoStop}%
\bibitem [{\citenamefont {Gekelman}\ \emph {et~al.}(2016)\citenamefont {Gekelman}, \citenamefont {Pribyl}, \citenamefont {Lucky}, \citenamefont {Drandell}, \citenamefont {Leneman}, \citenamefont {Maggs}, \citenamefont {Vincena}, \citenamefont {Compernolle}, \citenamefont {Tripathi}, \citenamefont {Morales}, \citenamefont {Carter}, \citenamefont {Wang},\ and\ \citenamefont {DeHaas}}]{Gekelman2016}%
  \BibitemOpen
  \bibfield  {author} {\bibinfo {author} {\bibfnamefont {W.}~\bibnamefont {Gekelman}}, \bibinfo {author} {\bibfnamefont {P.}~\bibnamefont {Pribyl}}, \bibinfo {author} {\bibfnamefont {Z.}~\bibnamefont {Lucky}}, \bibinfo {author} {\bibfnamefont {M.}~\bibnamefont {Drandell}}, \bibinfo {author} {\bibfnamefont {D.}~\bibnamefont {Leneman}}, \bibinfo {author} {\bibfnamefont {J.}~\bibnamefont {Maggs}}, \bibinfo {author} {\bibfnamefont {S.}~\bibnamefont {Vincena}}, \bibinfo {author} {\bibfnamefont {B.~V.}\ \bibnamefont {Compernolle}}, \bibinfo {author} {\bibfnamefont {S.~K.~P.}\ \bibnamefont {Tripathi}}, \bibinfo {author} {\bibfnamefont {G.}~\bibnamefont {Morales}}, \bibinfo {author} {\bibfnamefont {T.~A.}\ \bibnamefont {Carter}}, \bibinfo {author} {\bibfnamefont {Y.}~\bibnamefont {Wang}},\ and\ \bibinfo {author} {\bibfnamefont {T.}~\bibnamefont {DeHaas}},\ }\bibfield  {title} {\bibinfo {title} {The upgraded {L}arge {P}lasma {D}evice, a machine for studying frontier basic plasma physics},\ }\href
  {https://doi.org/10.1063/1.4941079} {\bibfield  {journal} {\bibinfo  {journal} {Rev. Sci. Instrum.}\ }\textbf {\bibinfo {volume} {87}},\ \bibinfo {pages} {025105} (\bibinfo {year} {2016})}\BibitemShut {NoStop}%
\bibitem [{\citenamefont {Qian}\ \emph {et~al.}(2023)\citenamefont {Qian}, \citenamefont {Gekelman}, \citenamefont {Pribyl}, \citenamefont {Sketchley}, \citenamefont {Tripathi}, \citenamefont {Lucky}, \citenamefont {Drandell}, \citenamefont {Vincena}, \citenamefont {Look}, \citenamefont {Travis}, \citenamefont {Carter}, \citenamefont {Wan}, \citenamefont {Cattelan}, \citenamefont {Sabiston}, \citenamefont {Ottaviano},\ and\ \citenamefont {Wirz}}]{Qian2023}%
  \BibitemOpen
  \bibfield  {author} {\bibinfo {author} {\bibfnamefont {Y.}~\bibnamefont {Qian}}, \bibinfo {author} {\bibfnamefont {W.}~\bibnamefont {Gekelman}}, \bibinfo {author} {\bibfnamefont {P.}~\bibnamefont {Pribyl}}, \bibinfo {author} {\bibfnamefont {T.}~\bibnamefont {Sketchley}}, \bibinfo {author} {\bibfnamefont {S.}~\bibnamefont {Tripathi}}, \bibinfo {author} {\bibfnamefont {Z.}~\bibnamefont {Lucky}}, \bibinfo {author} {\bibfnamefont {M.}~\bibnamefont {Drandell}}, \bibinfo {author} {\bibfnamefont {S.}~\bibnamefont {Vincena}}, \bibinfo {author} {\bibfnamefont {T.}~\bibnamefont {Look}}, \bibinfo {author} {\bibfnamefont {P.}~\bibnamefont {Travis}}, \bibinfo {author} {\bibfnamefont {T.}~\bibnamefont {Carter}}, \bibinfo {author} {\bibfnamefont {G.}~\bibnamefont {Wan}}, \bibinfo {author} {\bibfnamefont {M.}~\bibnamefont {Cattelan}}, \bibinfo {author} {\bibfnamefont {G.}~\bibnamefont {Sabiston}}, \bibinfo {author} {\bibfnamefont {A.}~\bibnamefont {Ottaviano}},\ and\ \bibinfo {author} {\bibfnamefont {R.}~\bibnamefont
  {Wirz}},\ }\bibfield  {title} {\bibinfo {title} {Design of the lanthanum hexaboride based plasma source for the large plasma device at {UCLA}},\ }\href {https://doi.org/10.1063/5.0152216} {\bibfield  {journal} {\bibinfo  {journal} {Rev. Sci. Instrum.}\ }\textbf {\bibinfo {volume} {94}},\ \bibinfo {pages} {085104} (\bibinfo {year} {2023})}\BibitemShut {NoStop}%
\bibitem [{\citenamefont {Morales}\ and\ \citenamefont {Maggs}(1997)}]{Morales1997}%
  \BibitemOpen
  \bibfield  {author} {\bibinfo {author} {\bibfnamefont {G.~J.}\ \bibnamefont {Morales}}\ and\ \bibinfo {author} {\bibfnamefont {J.~E.}\ \bibnamefont {Maggs}},\ }\bibfield  {title} {\bibinfo {title} {Structure of kinetic {A}lfvén waves with small transverse scale length},\ }\href {https://doi.org/10.1063/1.872531} {\bibfield  {journal} {\bibinfo  {journal} {Phys. Plasmas}\ }\textbf {\bibinfo {volume} {4}},\ \bibinfo {pages} {4118} (\bibinfo {year} {1997})}\BibitemShut {NoStop}%
\bibitem [{\citenamefont {Leneman}\ \emph {et~al.}(1999)\citenamefont {Leneman}, \citenamefont {Gekelman},\ and\ \citenamefont {Maggs}}]{Leneman1999}%
  \BibitemOpen
  \bibfield  {author} {\bibinfo {author} {\bibfnamefont {D.}~\bibnamefont {Leneman}}, \bibinfo {author} {\bibfnamefont {W.}~\bibnamefont {Gekelman}},\ and\ \bibinfo {author} {\bibfnamefont {J.}~\bibnamefont {Maggs}},\ }\bibfield  {title} {\bibinfo {title} {Laboratory observations of shear {A}lfvén waves launched from a small source},\ }\href {https://doi.org/10.1103/physrevlett.82.2673} {\bibfield  {journal} {\bibinfo  {journal} {Phys. Rev. Lett.}\ }\textbf {\bibinfo {volume} {82}},\ \bibinfo {pages} {2673} (\bibinfo {year} {1999})}\BibitemShut {NoStop}%
\bibitem [{\citenamefont {Gekelman}\ \emph {et~al.}(2011)\citenamefont {Gekelman}, \citenamefont {Vincena}, \citenamefont {Van~Compernolle}, \citenamefont {Morales}, \citenamefont {Maggs}, \citenamefont {Pribyl},\ and\ \citenamefont {Carter}}]{Gekelman2011}%
  \BibitemOpen
  \bibfield  {author} {\bibinfo {author} {\bibfnamefont {W.}~\bibnamefont {Gekelman}}, \bibinfo {author} {\bibfnamefont {S.}~\bibnamefont {Vincena}}, \bibinfo {author} {\bibfnamefont {B.}~\bibnamefont {Van~Compernolle}}, \bibinfo {author} {\bibfnamefont {G.~J.}\ \bibnamefont {Morales}}, \bibinfo {author} {\bibfnamefont {J.~E.}\ \bibnamefont {Maggs}}, \bibinfo {author} {\bibfnamefont {P.}~\bibnamefont {Pribyl}},\ and\ \bibinfo {author} {\bibfnamefont {T.~A.}\ \bibnamefont {Carter}},\ }\bibfield  {title} {\bibinfo {title} {The many faces of shear {A}lfvén waves},\ }\href {https://doi.org/10.1063/1.3592210} {\bibfield  {journal} {\bibinfo  {journal} {Phys. Plasmas}\ }\textbf {\bibinfo {volume} {18}},\ \bibinfo {pages} {055501} (\bibinfo {year} {2011})}\BibitemShut {NoStop}%
\bibitem [{\citenamefont {Gueroult}\ \emph {et~al.}(2024)\citenamefont {Gueroult}, \citenamefont {Tripathi}, \citenamefont {Gaboriau}, \citenamefont {Look},\ and\ \citenamefont {Fisch}}]{Gueroult2024}%
  \BibitemOpen
  \bibfield  {author} {\bibinfo {author} {\bibfnamefont {R.}~\bibnamefont {Gueroult}}, \bibinfo {author} {\bibfnamefont {S.}~\bibnamefont {Tripathi}}, \bibinfo {author} {\bibfnamefont {F.}~\bibnamefont {Gaboriau}}, \bibinfo {author} {\bibfnamefont {T.}~\bibnamefont {Look}},\ and\ \bibinfo {author} {\bibfnamefont {N.}~\bibnamefont {Fisch}},\ }\bibfield  {title} {\bibinfo {title} {Plasma potential shaping using end-electrodes in the large plasma device},\ }\href {https://doi.org/10.1017/S0022377824000552} {\bibfield  {journal} {\bibinfo  {journal} {J. Plasma Phys.}\ }\textbf {\bibinfo {volume} {90}},\ \bibinfo {pages} {905900603} (\bibinfo {year} {2024})}\BibitemShut {NoStop}%
\bibitem [{\citenamefont {Everson}\ \emph {et~al.}(2009)\citenamefont {Everson}, \citenamefont {Pribyl}, \citenamefont {Constantin}, \citenamefont {Zylstra}, \citenamefont {Schaeffer}, \citenamefont {Kugland},\ and\ \citenamefont {Niemann}}]{Everson2009}%
  \BibitemOpen
  \bibfield  {author} {\bibinfo {author} {\bibfnamefont {E.~T.}\ \bibnamefont {Everson}}, \bibinfo {author} {\bibfnamefont {P.}~\bibnamefont {Pribyl}}, \bibinfo {author} {\bibfnamefont {C.~G.}\ \bibnamefont {Constantin}}, \bibinfo {author} {\bibfnamefont {A.}~\bibnamefont {Zylstra}}, \bibinfo {author} {\bibfnamefont {D.}~\bibnamefont {Schaeffer}}, \bibinfo {author} {\bibfnamefont {N.~L.}\ \bibnamefont {Kugland}},\ and\ \bibinfo {author} {\bibfnamefont {C.}~\bibnamefont {Niemann}},\ }\bibfield  {title} {\bibinfo {title} {Design, construction, and calibration of a three-axis, high-frequency magnetic probe (b-dot probe) as a diagnostic for exploding plasmas},\ }\href {https://doi.org/10.1063/1.3246785} {\bibfield  {journal} {\bibinfo  {journal} {Rev. Sci. Instrum.}\ }\textbf {\bibinfo {volume} {80}},\ \bibinfo {pages} {113505} (\bibinfo {year} {2009})}\BibitemShut {NoStop}%
\bibitem [{Sup()}]{SupplementaryMaterial}%
  \BibitemOpen
  \href@noop {} {}\bibinfo {note} {See Supplemental Material at [URL will be inserted by publisher] for additional elements on the experimental results.}\BibitemShut {Stop}%
\bibitem [{\citenamefont {Rahbarnia}\ \emph {et~al.}(2010)\citenamefont {Rahbarnia}, \citenamefont {Ullrich}, \citenamefont {Sauer}, \citenamefont {Grulke},\ and\ \citenamefont {Klinger}}]{Rahbarnia2010}%
  \BibitemOpen
  \bibfield  {author} {\bibinfo {author} {\bibfnamefont {K.}~\bibnamefont {Rahbarnia}}, \bibinfo {author} {\bibfnamefont {S.}~\bibnamefont {Ullrich}}, \bibinfo {author} {\bibfnamefont {K.}~\bibnamefont {Sauer}}, \bibinfo {author} {\bibfnamefont {O.}~\bibnamefont {Grulke}},\ and\ \bibinfo {author} {\bibfnamefont {T.}~\bibnamefont {Klinger}},\ }\bibfield  {title} {\bibinfo {title} {Alfvén wave dispersion behavior in single- and multicomponent plasmas},\ }\href {https://doi.org/10.1063/1.3322852} {\bibfield  {journal} {\bibinfo  {journal} {Phys. Plasmas}\ }\textbf {\bibinfo {volume} {17}},\ \bibinfo {pages} {032102} (\bibinfo {year} {2010})}\BibitemShut {NoStop}%
\bibitem [{\citenamefont {Acheson}\ and\ \citenamefont {Hide}(1973)}]{Acheson1973}%
  \BibitemOpen
  \bibfield  {author} {\bibinfo {author} {\bibfnamefont {D.~J.}\ \bibnamefont {Acheson}}\ and\ \bibinfo {author} {\bibfnamefont {R.}~\bibnamefont {Hide}},\ }\bibfield  {title} {\bibinfo {title} {Hydromagnetics of rotating fluids},\ }\href {https://doi.org/10.1088/0034-4885/36/2/002} {\bibfield  {journal} {\bibinfo  {journal} {Rep. Progr. Phys.}\ }\textbf {\bibinfo {volume} {36}},\ \bibinfo {pages} {159} (\bibinfo {year} {1973})}\BibitemShut {NoStop}%
\bibitem [{\citenamefont {Rax}\ \emph {et~al.}(2023)\citenamefont {Rax}, \citenamefont {Gueroult},\ and\ \citenamefont {Fisch}}]{Rax2023b}%
  \BibitemOpen
  \bibfield  {author} {\bibinfo {author} {\bibfnamefont {J.-M.}\ \bibnamefont {Rax}}, \bibinfo {author} {\bibfnamefont {R.}~\bibnamefont {Gueroult}},\ and\ \bibinfo {author} {\bibfnamefont {N.}~\bibnamefont {Fisch}},\ }\bibfield  {title} {\bibinfo {title} {Rotating {A}lfvén waves in rotating plasmas},\ }\href {https://doi.org/10.1017/s0022377823001368} {\bibfield  {journal} {\bibinfo  {journal} {J. Plasma Phys.}\ }\textbf {\bibinfo {volume} {89}},\ \bibinfo {pages} {905890613} (\bibinfo {year} {2023})}\BibitemShut {NoStop}%
\bibitem [{\citenamefont {Pfeifer}\ \emph {et~al.}(2007)\citenamefont {Pfeifer}, \citenamefont {Nieminen}, \citenamefont {Heckenberg},\ and\ \citenamefont {Rubinsztein-Dunlop}}]{Pfeifer2007}%
  \BibitemOpen
  \bibfield  {author} {\bibinfo {author} {\bibfnamefont {R.~N.~C.}\ \bibnamefont {Pfeifer}}, \bibinfo {author} {\bibfnamefont {T.~A.}\ \bibnamefont {Nieminen}}, \bibinfo {author} {\bibfnamefont {N.~R.}\ \bibnamefont {Heckenberg}},\ and\ \bibinfo {author} {\bibfnamefont {H.}~\bibnamefont {Rubinsztein-Dunlop}},\ }\bibfield  {title} {\bibinfo {title} {Colloquium: Momentum of an electromagnetic wave in dielectric media},\ }\href {https://doi.org/10.1103/revmodphys.79.1197} {\bibfield  {journal} {\bibinfo  {journal} {Rev. Modern Phys.}\ }\textbf {\bibinfo {volume} {79}},\ \bibinfo {pages} {1197} (\bibinfo {year} {2007})}\BibitemShut {NoStop}%
\bibitem [{\citenamefont {Barnett}\ and\ \citenamefont {Loudon}(2010)}]{Barnett2010}%
  \BibitemOpen
  \bibfield  {author} {\bibinfo {author} {\bibfnamefont {S.~M.}\ \bibnamefont {Barnett}}\ and\ \bibinfo {author} {\bibfnamefont {R.}~\bibnamefont {Loudon}},\ }\bibfield  {title} {\bibinfo {title} {The enigma of optical momentum in a medium},\ }\href {https://doi.org/10.1098/rsta.2009.0207} {\bibfield  {journal} {\bibinfo  {journal} {Philos. Trans. R. Soc. A}\ }\textbf {\bibinfo {volume} {368}},\ \bibinfo {pages} {927} (\bibinfo {year} {2010})}\BibitemShut {NoStop}%
\bibitem [{\citenamefont {Bliokh}\ \emph {et~al.}(2017)\citenamefont {Bliokh}, \citenamefont {Bekshaev},\ and\ \citenamefont {Nori}}]{Bliokh2017}%
  \BibitemOpen
  \bibfield  {author} {\bibinfo {author} {\bibfnamefont {K.~Y.}\ \bibnamefont {Bliokh}}, \bibinfo {author} {\bibfnamefont {A.~Y.}\ \bibnamefont {Bekshaev}},\ and\ \bibinfo {author} {\bibfnamefont {F.}~\bibnamefont {Nori}},\ }\bibfield  {title} {\bibinfo {title} {Optical momentum, spin, and angular momentum in dispersive media},\ }\href {https://doi.org/10.1103/physrevlett.119.073901} {\bibfield  {journal} {\bibinfo  {journal} {Phys. Rev. Lett.}\ }\textbf {\bibinfo {volume} {119}},\ \bibinfo {pages} {073901} (\bibinfo {year} {2017})}\BibitemShut {NoStop}%
\bibitem [{\citenamefont {Lavery}\ \emph {et~al.}(2013)\citenamefont {Lavery}, \citenamefont {Speirits}, \citenamefont {Barnett},\ and\ \citenamefont {Padgett}}]{Lavery2013}%
  \BibitemOpen
  \bibfield  {author} {\bibinfo {author} {\bibfnamefont {M.~P.~J.}\ \bibnamefont {Lavery}}, \bibinfo {author} {\bibfnamefont {F.~C.}\ \bibnamefont {Speirits}}, \bibinfo {author} {\bibfnamefont {S.~M.}\ \bibnamefont {Barnett}},\ and\ \bibinfo {author} {\bibfnamefont {M.~J.}\ \bibnamefont {Padgett}},\ }\bibfield  {title} {\bibinfo {title} {Detection of a spinning object using light’s orbital angular momentum},\ }\href {https://doi.org/10.1126/science.1239936} {\bibfield  {journal} {\bibinfo  {journal} {Science}\ }\textbf {\bibinfo {volume} {341}},\ \bibinfo {pages} {537} (\bibinfo {year} {2013})}\BibitemShut {NoStop}%
\bibitem [{\citenamefont {Emile}\ \emph {et~al.}(2020)\citenamefont {Emile}, \citenamefont {Emile},\ and\ \citenamefont {Brousseau}}]{Emile2020}%
  \BibitemOpen
  \bibfield  {author} {\bibinfo {author} {\bibfnamefont {O.}~\bibnamefont {Emile}}, \bibinfo {author} {\bibfnamefont {J.}~\bibnamefont {Emile}},\ and\ \bibinfo {author} {\bibfnamefont {C.}~\bibnamefont {Brousseau}},\ }\bibfield  {title} {\bibinfo {title} {Rotational doppler shift upon reflection from a right angle prism},\ }\href {https://doi.org/10.1063/5.0009396} {\bibfield  {journal} {\bibinfo  {journal} {Appl. Phys. Lett.}\ }\textbf {\bibinfo {volume} {116}},\ \bibinfo {pages} {221102} (\bibinfo {year} {2020})}\BibitemShut {NoStop}%
\bibitem [{\citenamefont {Anderson}\ \emph {et~al.}(2022)\citenamefont {Anderson}, \citenamefont {Strong}, \citenamefont {Coburn}, \citenamefont {Rieker},\ and\ \citenamefont {Gopinath}}]{Anderson2022}%
  \BibitemOpen
  \bibfield  {author} {\bibinfo {author} {\bibfnamefont {A.~Q.}\ \bibnamefont {Anderson}}, \bibinfo {author} {\bibfnamefont {E.~F.}\ \bibnamefont {Strong}}, \bibinfo {author} {\bibfnamefont {S.~C.}\ \bibnamefont {Coburn}}, \bibinfo {author} {\bibfnamefont {G.~B.}\ \bibnamefont {Rieker}},\ and\ \bibinfo {author} {\bibfnamefont {J.~T.}\ \bibnamefont {Gopinath}},\ }\bibfield  {title} {\bibinfo {title} {Orbital angular momentum-based dual-comb interferometer for ranging and rotation sensing},\ }\href {https://doi.org/10.1364/oe.457238} {\bibfield  {journal} {\bibinfo  {journal} {Opt. Express}\ }\textbf {\bibinfo {volume} {30}},\ \bibinfo {pages} {21195} (\bibinfo {year} {2022})}\BibitemShut {NoStop}%
\bibitem [{\citenamefont {Ren}\ \emph {et~al.}(2022)\citenamefont {Ren}, \citenamefont {Qiu}, \citenamefont {Liu},\ and\ \citenamefont {Liu}}]{Ren2022}%
  \BibitemOpen
  \bibfield  {author} {\bibinfo {author} {\bibfnamefont {Y.}~\bibnamefont {Ren}}, \bibinfo {author} {\bibfnamefont {S.}~\bibnamefont {Qiu}}, \bibinfo {author} {\bibfnamefont {T.}~\bibnamefont {Liu}},\ and\ \bibinfo {author} {\bibfnamefont {Z.}~\bibnamefont {Liu}},\ }\bibfield  {title} {\bibinfo {title} {Compound motion detection based on {OAM} interferometry},\ }\href {https://doi.org/10.1515/nanoph-2021-0622} {\bibfield  {journal} {\bibinfo  {journal} {Nanophotonics}\ }\textbf {\bibinfo {volume} {11}},\ \bibinfo {pages} {1127} (\bibinfo {year} {2022})}\BibitemShut {NoStop}%
\bibitem [{\citenamefont {Gray}\ \emph {et~al.}(2014)\citenamefont {Gray}, \citenamefont {Pisano}, \citenamefont {Maccalli},\ and\ \citenamefont {Schemmel}}]{Gray2014}%
  \BibitemOpen
  \bibfield  {author} {\bibinfo {author} {\bibfnamefont {M.~D.}\ \bibnamefont {Gray}}, \bibinfo {author} {\bibfnamefont {G.}~\bibnamefont {Pisano}}, \bibinfo {author} {\bibfnamefont {S.}~\bibnamefont {Maccalli}},\ and\ \bibinfo {author} {\bibfnamefont {P.}~\bibnamefont {Schemmel}},\ }\bibfield  {title} {\bibinfo {title} {Amplification of oam radiation by astrophysical masers},\ }\href {https://doi.org/10.1093/mnras/stu1947} {\bibfield  {journal} {\bibinfo  {journal} {Mon. Not. R. Astron Soc.}\ }\textbf {\bibinfo {volume} {445}},\ \bibinfo {pages} {4477} (\bibinfo {year} {2014})}\BibitemShut {NoStop}%
\bibitem [{\citenamefont {Sanchez}\ and\ \citenamefont {Oesch}(2011)}]{Sanchez2011}%
  \BibitemOpen
  \bibfield  {author} {\bibinfo {author} {\bibfnamefont {D.~J.}\ \bibnamefont {Sanchez}}\ and\ \bibinfo {author} {\bibfnamefont {D.~W.}\ \bibnamefont {Oesch}},\ }\bibfield  {title} {\bibinfo {title} {Orbital angular momentum in optical waves propagating through distributed turbulence},\ }\href {https://doi.org/10.1364/oe.19.024596} {\bibfield  {journal} {\bibinfo  {journal} {Opt. Express}\ }\textbf {\bibinfo {volume} {19}},\ \bibinfo {pages} {24596} (\bibinfo {year} {2011})}\BibitemShut {NoStop}%
\bibitem [{\citenamefont {Katoh}\ \emph {et~al.}(2017)\citenamefont {Katoh}, \citenamefont {Fujimoto}, \citenamefont {Kawaguchi}, \citenamefont {Tsuchiya}, \citenamefont {Ohmi}, \citenamefont {Kaneyasu}, \citenamefont {Taira}, \citenamefont {Hosaka}, \citenamefont {Mochihashi},\ and\ \citenamefont {Takashima}}]{Katoh2017}%
  \BibitemOpen
  \bibfield  {author} {\bibinfo {author} {\bibfnamefont {M.}~\bibnamefont {Katoh}}, \bibinfo {author} {\bibfnamefont {M.}~\bibnamefont {Fujimoto}}, \bibinfo {author} {\bibfnamefont {H.}~\bibnamefont {Kawaguchi}}, \bibinfo {author} {\bibfnamefont {K.}~\bibnamefont {Tsuchiya}}, \bibinfo {author} {\bibfnamefont {K.}~\bibnamefont {Ohmi}}, \bibinfo {author} {\bibfnamefont {T.}~\bibnamefont {Kaneyasu}}, \bibinfo {author} {\bibfnamefont {Y.}~\bibnamefont {Taira}}, \bibinfo {author} {\bibfnamefont {M.}~\bibnamefont {Hosaka}}, \bibinfo {author} {\bibfnamefont {A.}~\bibnamefont {Mochihashi}},\ and\ \bibinfo {author} {\bibfnamefont {Y.}~\bibnamefont {Takashima}},\ }\bibfield  {title} {\bibinfo {title} {Angular momentum of twisted radiation from an electron in spiral motion},\ }\href {https://doi.org/10.1103/physrevlett.118.094801} {\bibfield  {journal} {\bibinfo  {journal} {Phys. Rev. Lett.}\ }\textbf {\bibinfo {volume} {118}},\ \bibinfo {pages} {094801} (\bibinfo {year} {2017})}\BibitemShut {NoStop}%
\bibitem [{\citenamefont {Harwit}(2003)}]{Harwit2003}%
  \BibitemOpen
  \bibfield  {author} {\bibinfo {author} {\bibfnamefont {M.}~\bibnamefont {Harwit}},\ }\bibfield  {title} {\bibinfo {title} {Photon orbital angular momentum in astrophysics},\ }\href {https://doi.org/10.1086/378623} {\bibfield  {journal} {\bibinfo  {journal} {Astrophys. J.}\ }\textbf {\bibinfo {volume} {597}},\ \bibinfo {pages} {1266} (\bibinfo {year} {2003})}\BibitemShut {NoStop}%
\bibitem [{\citenamefont {Thidé}\ \emph {et~al.}(2007)\citenamefont {Thidé}, \citenamefont {Then}, \citenamefont {Sjöholm}, \citenamefont {Palmer}, \citenamefont {Bergman}, \citenamefont {Carozzi}, \citenamefont {Istomin}, \citenamefont {Ibragimov},\ and\ \citenamefont {Khamitova}}]{Thide2007}%
  \BibitemOpen
  \bibfield  {author} {\bibinfo {author} {\bibfnamefont {B.}~\bibnamefont {Thidé}}, \bibinfo {author} {\bibfnamefont {H.}~\bibnamefont {Then}}, \bibinfo {author} {\bibfnamefont {J.}~\bibnamefont {Sjöholm}}, \bibinfo {author} {\bibfnamefont {K.}~\bibnamefont {Palmer}}, \bibinfo {author} {\bibfnamefont {J.}~\bibnamefont {Bergman}}, \bibinfo {author} {\bibfnamefont {T.~D.}\ \bibnamefont {Carozzi}}, \bibinfo {author} {\bibfnamefont {Y.~N.}\ \bibnamefont {Istomin}}, \bibinfo {author} {\bibfnamefont {N.~H.}\ \bibnamefont {Ibragimov}},\ and\ \bibinfo {author} {\bibfnamefont {R.}~\bibnamefont {Khamitova}},\ }\bibfield  {title} {\bibinfo {title} {Utilization of photon orbital angular momentum in the low-frequency radio domain},\ }\href {https://doi.org/10.1103/physrevlett.99.087701} {\bibfield  {journal} {\bibinfo  {journal} {Phys. Rev. Lett.}\ }\textbf {\bibinfo {volume} {99}},\ \bibinfo {pages} {087701} (\bibinfo {year} {2007})}\BibitemShut {NoStop}%
\bibitem [{\citenamefont {Elias}(2008)}]{Elias2008}%
  \BibitemOpen
  \bibfield  {author} {\bibinfo {author} {\bibfnamefont {N.~M.}\ \bibnamefont {Elias}},\ }\bibfield  {title} {\bibinfo {title} {Photon orbital angular momentum in astronomy},\ }\href {https://doi.org/10.1051/0004-6361:200809791} {\bibfield  {journal} {\bibinfo  {journal} {Astron. \& Astrophys.}\ }\textbf {\bibinfo {volume} {492}},\ \bibinfo {pages} {883} (\bibinfo {year} {2008})}\BibitemShut {NoStop}%
\bibitem [{\citenamefont {Anzolin}\ \emph {et~al.}(2008)\citenamefont {Anzolin}, \citenamefont {Tamburini}, \citenamefont {Bianchini}, \citenamefont {Umbriaco},\ and\ \citenamefont {Barbieri}}]{Anzolin2008}%
  \BibitemOpen
  \bibfield  {author} {\bibinfo {author} {\bibfnamefont {G.}~\bibnamefont {Anzolin}}, \bibinfo {author} {\bibfnamefont {F.}~\bibnamefont {Tamburini}}, \bibinfo {author} {\bibfnamefont {A.}~\bibnamefont {Bianchini}}, \bibinfo {author} {\bibfnamefont {G.}~\bibnamefont {Umbriaco}},\ and\ \bibinfo {author} {\bibfnamefont {C.}~\bibnamefont {Barbieri}},\ }\bibfield  {title} {\bibinfo {title} {Optical vortices with starlight},\ }\href {https://doi.org/10.1051/0004-6361:200810469} {\bibfield  {journal} {\bibinfo  {journal} {Astron. \& Astrophys.}\ }\textbf {\bibinfo {volume} {488}},\ \bibinfo {pages} {1159} (\bibinfo {year} {2008})}\BibitemShut {NoStop}%
\bibitem [{\citenamefont {Gueroult}\ \emph {et~al.}(2019)\citenamefont {Gueroult}, \citenamefont {Shi}, \citenamefont {Rax},\ and\ \citenamefont {Fisch}}]{Gueroult2019}%
  \BibitemOpen
  \bibfield  {author} {\bibinfo {author} {\bibfnamefont {R.}~\bibnamefont {Gueroult}}, \bibinfo {author} {\bibfnamefont {Y.}~\bibnamefont {Shi}}, \bibinfo {author} {\bibfnamefont {J.-M.}\ \bibnamefont {Rax}},\ and\ \bibinfo {author} {\bibfnamefont {N.~J.}\ \bibnamefont {Fisch}},\ }\bibfield  {title} {\bibinfo {title} {Determining the rotation direction in pulsars},\ }\href {https://doi.org/10.1038/s41467-019-11243-4} {\bibfield  {journal} {\bibinfo  {journal} {Nat. Commun.}\ }\textbf {\bibinfo {volume} {10}},\ \bibinfo {pages} {3232} (\bibinfo {year} {2019})}\BibitemShut {NoStop}%
\bibitem [{\citenamefont {Lehnert}(1971)}]{Lehnert1971}%
  \BibitemOpen
  \bibfield  {author} {\bibinfo {author} {\bibfnamefont {B.}~\bibnamefont {Lehnert}},\ }\bibfield  {title} {\bibinfo {title} {Rotating plasmas},\ }\href {https://doi.org/10.1088/0029-5515/11/5/010} {\bibfield  {journal} {\bibinfo  {journal} {Nucl. Fusion}\ }\textbf {\bibinfo {volume} {11}},\ \bibinfo {pages} {485} (\bibinfo {year} {1971})}\BibitemShut {NoStop}%
\bibitem [{\citenamefont {Fisch}\ and\ \citenamefont {Rax}(1992)}]{Fisch1992}%
  \BibitemOpen
  \bibfield  {author} {\bibinfo {author} {\bibfnamefont {N.~J.}\ \bibnamefont {Fisch}}\ and\ \bibinfo {author} {\bibfnamefont {J.-M.}\ \bibnamefont {Rax}},\ }\bibfield  {title} {\bibinfo {title} {Interaction of energetic alpha particles with intense lower hybrid waves},\ }\href {https://doi.org/10.1103/PhysRevLett.69.612} {\bibfield  {journal} {\bibinfo  {journal} {Phys. Rev. Lett.}\ }\textbf {\bibinfo {volume} {69}},\ \bibinfo {pages} {612} (\bibinfo {year} {1992})}\BibitemShut {NoStop}%
\bibitem [{\citenamefont {Fetterman}\ and\ \citenamefont {Fisch}(2008)}]{Fetterman2008}%
  \BibitemOpen
  \bibfield  {author} {\bibinfo {author} {\bibfnamefont {A.~J.}\ \bibnamefont {Fetterman}}\ and\ \bibinfo {author} {\bibfnamefont {N.~J.}\ \bibnamefont {Fisch}},\ }\bibfield  {title} {\bibinfo {title} {$\alpha$ channeling in a rotating plasma},\ }\href@noop {} {\bibfield  {journal} {\bibinfo  {journal} {Phys. Rev. Lett.}\ }\textbf {\bibinfo {volume} {101}},\ \bibinfo {pages} {205003} (\bibinfo {year} {2008})}\BibitemShut {NoStop}%
\bibitem [{\citenamefont {Fetterman}\ and\ \citenamefont {Fisch}(2010)}]{Fetterman2010}%
  \BibitemOpen
  \bibfield  {author} {\bibinfo {author} {\bibfnamefont {A.~J.}\ \bibnamefont {Fetterman}}\ and\ \bibinfo {author} {\bibfnamefont {N.~J.}\ \bibnamefont {Fisch}},\ }\bibfield  {title} {\bibinfo {title} {Alpha channeling in rotating plasma with stationary waves},\ }\href {https://doi.org/10.1063/1.3389308} {\bibfield  {journal} {\bibinfo  {journal} {Phys. Plasmas}\ }\textbf {\bibinfo {volume} {17}},\ \bibinfo {pages} {042112} (\bibinfo {year} {2010})}\BibitemShut {NoStop}%
\bibitem [{\citenamefont {Kolmes}\ \emph {et~al.}(2024)\citenamefont {Kolmes}, \citenamefont {Ochs}, \citenamefont {Rax},\ and\ \citenamefont {Fisch}}]{Kolmes2024VoltageDrop}%
  \BibitemOpen
  \bibfield  {author} {\bibinfo {author} {\bibfnamefont {E.~J.}\ \bibnamefont {Kolmes}}, \bibinfo {author} {\bibfnamefont {I.~E.}\ \bibnamefont {Ochs}}, \bibinfo {author} {\bibfnamefont {J.-M.}\ \bibnamefont {Rax}},\ and\ \bibinfo {author} {\bibfnamefont {N.~J.}\ \bibnamefont {Fisch}},\ }\bibfield  {title} {\bibinfo {title} {Massive, long-lived electrostatic potentials in a rotating mirror plasma},\ }\href {https://doi.org/10.1038/s41467-024-47386-2} {\bibfield  {journal} {\bibinfo  {journal} {Nat. Commun.}\ }\textbf {\bibinfo {volume} {15}},\ \bibinfo {pages} {4302} (\bibinfo {year} {2024})}\BibitemShut {NoStop}%
\bibitem [{\citenamefont {Gueroult}(2025)}]{Zenodo2025}%
  \BibitemOpen
  \bibfield  {author} {\bibinfo {author} {\bibfnamefont {R.}~\bibnamefont {Gueroult}},\ }\bibfield  {title} {\bibinfo {title} {Dataset for "{I}mage rotation in plasmas"},\ }\href {https://doi.org/10.5281/zenodo.15467967} {10.5281/zenodo.15467967} (\bibinfo {year} {2025})\BibitemShut {NoStop}%
\end{thebibliography}
%

\section*{End matter}

\emph{Aberration for low-frequency waves--}
Writing $\Sigma$ and $\Sigma'$ the laboratory frame and the rest-frame, and denoting again with a prime quantities expressed in $\Sigma'$, the Lorentz-transformation for the wave four-vector ($\omega, \mathbf{k}$) gives to lowest order in $v/c$ an aberration 
\begin{equation}
\mathbf{k}' = \mathbf{k} - \mathbf{v}\omega/c^2.
\end{equation}
For the wave frequency $\omega_0\sim10^6$~s$^{-1}$ studied experimentally and the velocities $v$ of a few km.s$^{-1}$ achieved in LAPD, one finds a wavevector correction $|\mathbf{v}\omega/c^2|\leq10^{-7}~$m$^{-1}$. These corrections are negligible for the centimeter scale Alfv{\'e}n waves studied here. As a result the rest-frame wavevector $\mathbf{k}'$ can in first approximation be taken equal to the lab-frame wavevector $\mathbf{k}$.

\emph{Kinetic shear Alfv{\'e}n wave--} The simple dispersion relation Eq.~\eqref{Eq:disp_Alf} used in the main text to infer a group velocity $\mathbf{v_g'}=v_A\mathbf{\hat{e}}_z$ is only valid for a plane wave propagating exactly parallel to the confinement magnetic field $\mathbf{k}' = k'\mathbf{\hat{e}}_z$, and in the limit $\omega'\ll\Omega_{ci}'$ with $\Omega_{ci}'$ the rest-frame ion cyclotron frequency. As we will now show both hypotheses need to be reconsidered for the LAPD experiment of interest here. First, for the operating conditions given in Table~\ref{Tab:conditions}, $\omega_0/\Omega_{ci}=0.4$. For non-relativistic velocities $\Omega_{ci}'\sim\Omega_{ci}$, whereas for $v\leq3$~km.s$^{-1}$ the Doppler shift $k_{\perp}v$ accounts for at most $20\%$ variation in wave frequency. Putting these pieces together we find $\omega_0'/\Omega_{ci}'=0.4\pm0.08$, confirming that the low frequency assumption is indeed not strictly verified. Second, taking the perpendicular wavelength equal to the antenna diameter, $k_{\perp}\sim60$~ m$^{-1}$, so that $k_{\perp}/k_{\parallel}\sim30$. This confirms that aberration effects are negligible, and from there that $k_{\perp}'$ is finite.

For oblique propagation in a finite temperature plasma, propagative solutions are generally complex and difficult to expose. Yet, simpler solutions can be elicited when considering the limit of waves with small transverse scalelength. Specifically, if the perpendicular wavevector is small compared to the ion Larmor radius, and if the Alfv{\'e}n velocity is smaller than the electron thermal speed $v_{the}=(k T_e/m_e)^{1/2}$, plane wave solutions take the form of kinetic shear Alfv{\'e}n waves (KSAW) with the dispersion relation~\cite{Leneman1999,Gekelman2011}
\begin{equation}
\label{Eq:disp_KSAW}
\frac{{\omega'}^2}{{{k}_{\parallel}'}^2}={v_A'}^2\left[1-\left(\frac{\omega'}{\Omega_{ci}'}\right)^2+{{k}_{\perp}'}^2{{\rho_s}'}^2\right]
\end{equation}
where $\rho_s'=c_s'/\Omega_{ci}'$ is the ion gyroradius, with $c_s'=(k T_e/m_i')^{1/2}$ the ion sound speed. Neglecting relativistic corrections, we drop the prime on $\Omega_{ci}$, $v_A$ and $\rho_s$, and neglect aberration $\mathbf{k}'=\mathbf{k}$. For the parameters given in Tab.~\ref{Tab:conditions} and $T_e=5$~eV, one finds $v_{the}\sim2.7v_A$ and $k_{\perp}\rho_s=0.4$, confirming the validity of these hypotheses in our experiment. Also, since $\omega_0/\Omega_{ci}\sim k_{\perp}\rho_s=0.4$ for the $154$~kHz wave frequency used here, the last two terms in brackets in Eq.~\eqref{Eq:disp_KSAW} approximately cancel each other. The parallel rest-frame phase velocity is thus approximately $v_A$, with the Doppler shift accounting for variations of at most $4\%$.

Deriving Eq.~\eqref{Eq:disp_KSAW} with respect to the rest-frame wavector $\mathbf{k}'$ gives the parallel and perpendicular rest-frame group velocity
\begin{equation}
v'_{g_{\parallel}}=\frac{\partial \omega'}{\partial k'_{\parallel}} = v_A \left[\frac{1+{k'_{\perp}}^2{\rho_s}^2}{\left(1+{k_{\parallel}'}^2 {v_A}^2/{\Omega_{ci}}^2\right)^{3}}\right]^{1/2}.
\end{equation}
and
\begin{equation}
v'_{g_{\perp}}=\frac{\partial \omega'}{\partial k'_{\perp}} = v_A \frac{k_{\parallel}'}{k_{\perp}'}\frac{{k'_{\perp}}^2{\rho_s}^2}{\left[(1+{k'_{\perp}}^2{\rho_s}^2)(1+{k_{\parallel}'}^2 {v_A}^2/{\Omega_{ci}}^2)\right]^{1/2}}.
\end{equation}
Quantifying the different terms in our experiment, one finds that the parallel component is slightly reduced compared to the simple Alfv{\'e}n case with $v'_{g_{\parallel}}\sim 0.86 v_A$. Also, the perpendicular component is now finite, but remains much smaller with $v'_{g_{\perp}}/v_A\leq 1/200$. This shows that the rest-frame group velocity of the KSAW remains largely along the magnetic field~\cite{Gekelman2011}. Note that for a small localized source this perpendicular component is directed radially outward, manifesting as Alfv{\'e}n wave cones~\cite{Morales1997}. 

\emph{Drag for a kinetic shear Alfv{\'e}n wave--} To characterize the drag, one needs the group velocity in the lab-frame. In general, this involves substituting in the dispersion relation Eq.~(\ref{Eq:disp_KSAW}) the Lorentz transformed wave four-vector ($\omega, \mathbf{k}$), and deriving the solution with respect to $\mathbf{k}$, instead of $\mathbf{k}'$. However, because aberration effects are here negligible, we can instead use a simpler route, and simply consider the transformation of the group velocity. Because the group velocity is small compared to the speed of light, we only consider correction to first order in $\beta=v/c$,
\begin{equation}
\mathbf{v_g} =  \mathbf{v_g'}(1-c^{-2}\mathbf{v_g'}\cdot\mathbf{v}) +  \mathbf{v}. 
\end{equation}
For a motion entirely in the azimuthal direction, and a rest frame group velocity that is zero in this direction $\mathbf{v_g'}\cdot\mathbf{\hat{e}}_{\theta}=0$, we find that only the azimuthal component of the group velocity is corrected by the medium's velocity $v$, with $v_{g_{\theta}}\sim v$. To first order in $\beta$ the associated beam deviation angle is thus
\begin{equation}
\frac{v_{g_{\theta}}}{v_{g_{\parallel}}} \sim \frac{v}{v'_{g_{\parallel}}},
\end{equation}
leading to a lateral displacement
\begin{equation}
d\sim lv/v'_{g_{\parallel}}
\end{equation}
over a distance $l$.

\begin{figure}[htbp]
\begin{center}
\includegraphics[width=8.6cm]{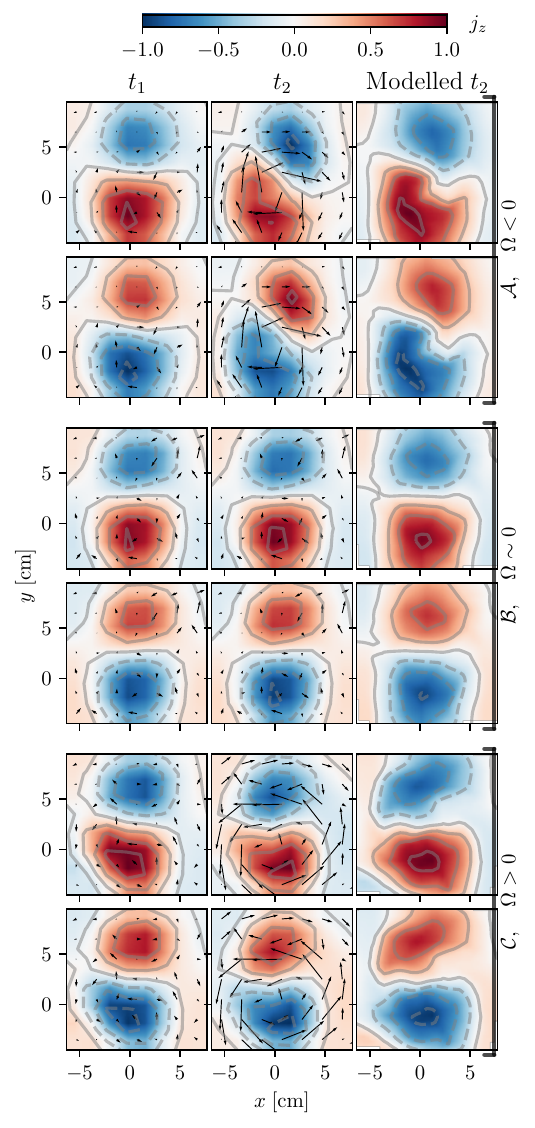}
\caption{Map of the normalized axial current $j_{z}$ in the $(xy)$ plane around the antenna center $(x_{a},y_{a}) = (0.7,2.5)$~cm on port $\#32$ at two opposite phases during a cycle before biasing ($t_{1}$, leftmost column), during active biasing ($t_{2}$, middle column) and reconstructed from theory (rightmost column). For the two leftmost columns the overlaid arrows show the velocity field inferred from the floating potential on port $\#34$ at the same instants. The three vertical groups correspond to the three biasing scenarios $\mathcal{A}$, $\mathcal{B}$ and $\mathcal{C}$. The transverse drag is modelled here with an increased group velocity $v_{g}=1.16v_{A}$ as compared to Fig.~\eqref{Fig:RotationDuringBiasing}, as expected for the kinetic shear Alfv{\'e}n wave.} 
\label{Fig:RotationDuringBiasing_1.5}
\end{center}
\end{figure}

This last result shows that that the azimuthal drag expected for the kinetic shear Alfv{\'e}n wave produced in LAPD is very much comparable to Eq~(\ref{Eq:drag}) obtained for the simple Alfv{\'e}n wave Eq.~\eqref{Eq:disp_Alf} propagating along the magnetic field. The main difference is an enhancement of drag effects by a factor $1/0.86\sim 1.16$ due to the slower parallel rest-frame group velocity. This is shown in Fig.~\ref{Fig:RotationDuringBiasing_1.5}, showing an improved match with the experimental data.

\emph{Image rotation--} Considering OAM carrying modes of the form
\begin{equation}
\Phi^{\pm} = J_{m}(\alpha r)\exp\left[i(k_{\parallel}^{\pm}z\pm m\theta-\omega t)\right],
\end{equation}
it was shown in Ref.~\cite{Rax2023b} that the rotation per unit length of the transverse structure of a shear Alfv{\'e}n wave obtained from the superposition of two modes with opposite azimuthal wave number $\pm m$ is 
\begin{equation}
\label{Eq:OAM_rot}
\varphi = \frac{1}{2}\frac{\Omega}{\omega}k_{\parallel}\mathcal{K}^{2} \frac{k_{\parallel}^{2}{v_{A}}^{2}+\omega^{2}}{k_{\parallel}^{2}{v_{A}}^{2}(\mathcal{K}^{2}+k_{\parallel}^{2})+\omega^{2}(\mathcal{K}^{2}-k_{\parallel}^{2})}
\end{equation}
with 
\begin{equation}
\alpha^{2}+k_{\parallel}^{2}=\mathcal{K}^{2}.
\end{equation}
Recalling that in our conditions $\omega/\Omega_{ci}\sim k_{\perp}\rho_s$, Eq.~\eqref{Eq:disp_KSAW} conveniently gives $\omega\sim k_{\parallel}v_{A}$, and Eq.~\eqref{Eq:OAM_rot} then simply reduces to
\begin{equation}
\label{Eq:OAM_rot2}
\varphi \sim \frac{1}{2}\frac{\Omega}{\omega}k_{\parallel}\sim \frac{1}{2}\frac{\Omega}{v_{A}}.
\end{equation}
\vspace{7cm}

\end{document}